\documentclass[10pt, twocolumn,epsf]{IEEEtran}
\usepackage{graphicx, epstopdf,cite,subfigure,cite}
\usepackage{amsthm,amsmath,amssymb,algorithm,algcompatible}
\usepackage{amsfonts}%
\usepackage[cmex10]{mathtools}%
\usepackage{multirow}
\usepackage{url}
\usepackage{graphics}
\usepackage{array}
\usepackage{color,soul}
\usepackage{url, cite}
\usepackage{bbold}
\usepackage{float}
\usepackage{array}
\usepackage{adjustbox}
\usepackage{graphics}

\newcommand*{\Scale}[2][4]{\scalebox{#1}{$#2$}}%

\newtheorem{theorem}{\bf Theorem}

\newtheorem{lemma}{{\bf Lemma}}

\begin{document}
\title {A Split-Central-Buffered Load-Balancing Clos-Network Switch with In-Order Forwarding}

\author{Oladele~Theophilus Sule, Roberto~Rojas-Cessa,~\IEEEmembership{Senior Member,~IEEE,} Ziqian Dong,~\IEEEmembership{Senior Member,~IEEE,} Chuan-Bi Lin,~\IEEEmembership{Member,~IEEE}
	\thanks{This paper is an extended version of that published in IEEE trans. on Networking. \textit{(Corresponding author: Oladele Theophilus Sule)}}
	\thanks{O.T. Sule and R. Rojas-Cessa are with the Department of Electrical and Computer Engineering, New Jersey Institute of Technology, Newark, NJ 07102. Email: \{{ots5, rojas}\}@njit.edu.}
	\thanks{Z. Dong is with the Department of Electrical and Computer Engineering, New York Institute of Technology, New York, NY 10023.}
	\thanks{C. Lin is with the Department of Information and Communication Engineering, Chaoyang University of Technology, Wufeng District, Taichung, 41349, Taiwan.}
	\thanks{This work was partially supported by National Science Foundation (NSF) under Grant No. (CNS) 1641033.}}

\date{January-21-2018}
\maketitle \thispagestyle{empty}

\begin{abstract}
We propose a configuration scheme for a load-balancing Clos-network packet switch that has split central modules and buffers in between the split modules. Our split-central-buffered Load-Balancing Clos-network (LBC) switch is cell based. The switch has four stages, namely input, central-input, central-output, and output stages. The proposed configuration scheme uses a pre-determined and periodic interconnection pattern in the input and split central modules to load-balance and route traffic. The LBC switch has low configuration complexity. The operation of the switch includes a mechanism applied at input and split-central modules to forward cells in sequence. 
The switch achieves 100\% throughput under uniform and nonuniform admissible traffic with independent and identical distributions (i.i.d.). These high switching performance and low complexity are achieved while performing in-sequence forwarding and without resorting to memory speedup or central-stage expansion. Our discussion includes throughput analysis, where we describe the operations that the configuration mechanism performs on the traffic traversing the switch, and proof of in-sequence forwarding. A simulation study is presented as a practical demonstration of the switch performance on uniform and nonuniform i.i.d. traffic.
\end{abstract}


\section{Introduction}
\label{lbc_intro}
Clos-network switches are attractive for building large-size switches \cite{clos1953study}. These switches mostly employ three stages, where each stage uses switch modules as building blocks. Each module is a small- or medium-size switch. Modules of the first, second, and third stages are often called input, central, and output modules, and they are denoted as IM, CM, and OM, respectively. Overall, Clos-network switches require fewer crosspoint elements, each of which is the atomic switching unit of a packet switch, than a single-stage switch of equivalent size, and thus they may require less building hardware. 
This trait of a Clos network often comes at the cost of an increased configuration complexity. The term configuration here means the local interconnection between inputs and outputs of a module.
 In general, a Clos-network switch requires the configuration of the modules in every stage before packets are forwarded through. 
Moreover, owing to the multi-stage architecture of such switch, the time for switch reconfiguration increases as the number of stages holding dependences increases. In a multi-stage switch, there is a dependence when the configuration of a module is affected by the configuration of another. The required configuration time dictates the internal data transmission time, which in turn defines the minimum size of the internal data unit. 
For example, switches that require long configuration time may need to use a long internal segment and time to transmit data while switches with fast configuration times may use a smaller segment size.
Therefore, the configuration time of a switch must be kept to the shortest possible for a fast and efficient reconfiguration \cite{al2008concatenating}. 

In the remainder of this paper, we consider the proposed packet switch to be cell based; that is, upon arrival at an input port of a switch, packets of variable size are segmented into fixed-size cells. Cells are forwarded through the switch to their destination outputs. Packets are re-assembled at the outputs of the switch. The selection of the cell length is left for the implementation of the LBC switch. However, as in any other switch, the cell length is decided by the time required to reconfigure IMs and CIMs and memory speed (of central queues or CBs). Cell length may be selected such that cell transmission time is equal to or greater than the largest of the switch configuration or memory response times. Additionally, the cell length can be increased if the average Internet packet is longer than the configuration time to reduce segmentation/reassembly processing \cite{al2008concatenating}.

Based on the design of its switching modules, each stage of a Clos-network switch can be categorized as
either space-based (S) or memory-based, where space switching modules are bufferless while memory switching modules are buffered. Space switching refers to the use of a level of parallelism where multiple cells can be switched at the same time slot by using multiple connections. Memory switching refers to the use of memory to store cells when they cannot be forwarded to the outputs (or next stage). Some of these categorizes are SSS (or S$^3$) \cite{lee1997path,chao2003matching}, MSM \cite{atlanta,kleban2013static,kleban2007modified,rojas2004maximum},  MMM \cite{chao2005trueway, chrysos2006scheduling, dong2011non, xia2016practical}, SMM \cite{li2005space}, and SSM \cite{lin2013minimizing,rojas2006scalable}, among the most popular ones. Out of those, S$^3$ switches require small amounts of hardware but their configuration has been proven challenging as input-to-output path setup must be resolved before cells are transmitted. On the other hand, inclusion of memory in modules may relax the configuration complexity. However, configuration complexity has remained high despite using memory in every switch module because of internal blocking and the multiplicity of input-output paths associated with diverse queuing delays \cite{chao2005trueway,dong2012mcs}. Specifically, switches with buffered central or output stages are prone to forwarding packets out of sequence, making re-sequencing or in-sequence transmission mechanisms an added feature. Moreover, the number and size of queues in a module are restricted to the available on-chip real estate. This restriction plus the adopted in-sequence measures may exacerbate internal blocking that, in turn, may lead to performance degradation \cite{dong2011non}.

Minimizing the complexity of the central module of a Clos-network switch has been of research interest in recent years. Hassen et. al proposed a Clos-network switch that combines different switching stages \cite{hassen2017high}. In this work, central modules are replaced with multi-directional networks-on-chip (MDN) modules. The switch uses a static dispatching scheme from the input/output modules, for which every input constantly delivers packets to the same MDN module, and adopts inter-central-module routing to enable forwarding of the cells to the final destination.
However, this switch may forward cells to the output ports out of sequence if cells from the same flow are routed through different paths on the central modules.

Load balancing traffic prior to routing it towards the destination output is a technique that not only improves switching performance but also reduces the configuration complexity of a packet switch when the load-balancing and routing follow a deterministic schedule \cite{chang2002load}. Such a schedule may be obtained as an application of matrix decomposition \cite{birkhoff1946tres, lee1996new}. This technique enables high performance not only on switches but also on a large number of network applications \cite{multipath-switching-slice12}.

A switch that load-balances traffic may need at least two stages to operate; one for load balancing and the other for routing cells to their destination outputs \cite{chang2002load}.  
A switch with such a deterministic and periodic schedule may require the use of queues between the load-balancing and routing stages. However, placing such queues and enabling multiple interconnection paths between an input and an output make load-balancing switches susceptible to forwarding cells out of sequence \cite{chang2002load}. This issue has been addressed by introducing either re-sequencing buffers at the output ports \cite{changload} or mechanisms that prevent out-of-sequence forwarding \cite{keslassy2002maintaining,rojas2016interconnections}.  However, these approaches are either complex or degrade switching performance.

Load balancing has been applied to Clos-network switches \cite{chao2005trueway, zhang2014space}. For example, Zhang et al. \cite{zhang2014space} proposed an SMM switch which adopts the two-stage load-balanced Birkhoff-von Neumann switch in each central module but has no input port buffers. Here, a central module consists of two $k \times k$ bufferless crossbar switches and $k$ buffers in between the crossbars. The switch performs load balancing at the input module and the first stage of the load-balanced Birkhoff-Von Neumann switch. 
Each of these queues accommodates up to one cell to guarantee the transmission of cells in sequence. However, the distance between modules in a large switch requires larger queue sizes for which this switch would suffer from out-of-sequence forwarding.

The switches discussed above suffer from either limited switching performance, high complexity, or out-of-sequence forwarding. These drawbacks then raise the question, can a load-balancing Clos-network switch achieve high switching performance, low configuration complexity, and in-sequence cell forwarding without resorting to memory speedup? 

In this paper, we aim at answering this question by proposing a split-central-buffered Load-Balancing Clos-network (LBC) switch. The switch has a split central module and queues in between. The switch employs predetermined and periodic interconnection patterns to interconnect the inputs and outputs of the switch modules. The switch load balances the incoming traffic and switches the cells towards the destination outputs, both with minimum configuration complexity. The result is a switch that attains high throughput under admissible traffic with independent and identical distribution (i.i.d.) and uses a configuration scheme with $O(1)$ complexity. The switch also adopts an in-sequence forwarding mechanism at the input queues to keep cells in sequence despite the presence of buffers between the split CMs. 

Different from existing switching architectures, as discussed above, the LBC switch achieves high performance, configuration simplicity, and in-sequence service, all attained without memory speedup nor central module expansion.

We analyze the performance of the proposed switch by modeling the effect of each stage on the traffic passing through the switch. In addition, we study the performance of the switch through traffic analysis and computer simulation. We show that the throughput of the switch approaches 100\% under several admissible traffic models, including traffic with nonuniform distributions, and demonstrate that the switch forwards cells to the output ports in sequence. The high performance and the in-sequence forwarding of packets of the switch are both achieved without resorting to speedup throughout the switch.

In summary, the contributions of this paper are as follows: 1) the proposal of a configuration scheme for a split-central-buffered load-balancing switch such that the attained throughput is 100\% under admissible traffic while having $O(1)$ scheduling complexity, 2) the proposal of an in-sequence mechanism for forwarding of cells in sequence throughout the switch, 3) the presentation of throughput analysis of the LBC switch for each of the stages that shows that the switch achieves 100\% throughput under i.i.d. admissible traffic, and 4) proof of the in-sequence capability of the proposed in-sequence forwarding mechanism.

The remainder of this paper is organized as follows: Section \ref{sec:switch-architecture} introduces the LBC switch. Section \ref{sec:throughput-analysis} analyzes the throughput performance of the proposed switch.  Section \ref{sec:proof} analyzes the in-sequence forwarding property of the LBC switch. Section \ref{sec:simulation-performance} presents a simulation study on the performance of the proposed switch. Section \ref{sec:conclusions} presents our conclusions. 

\section{Switch Architecture}
\label{sec:switch-architecture}

\begin{figure*}[htb]
 	\includegraphics[width=6.5in]{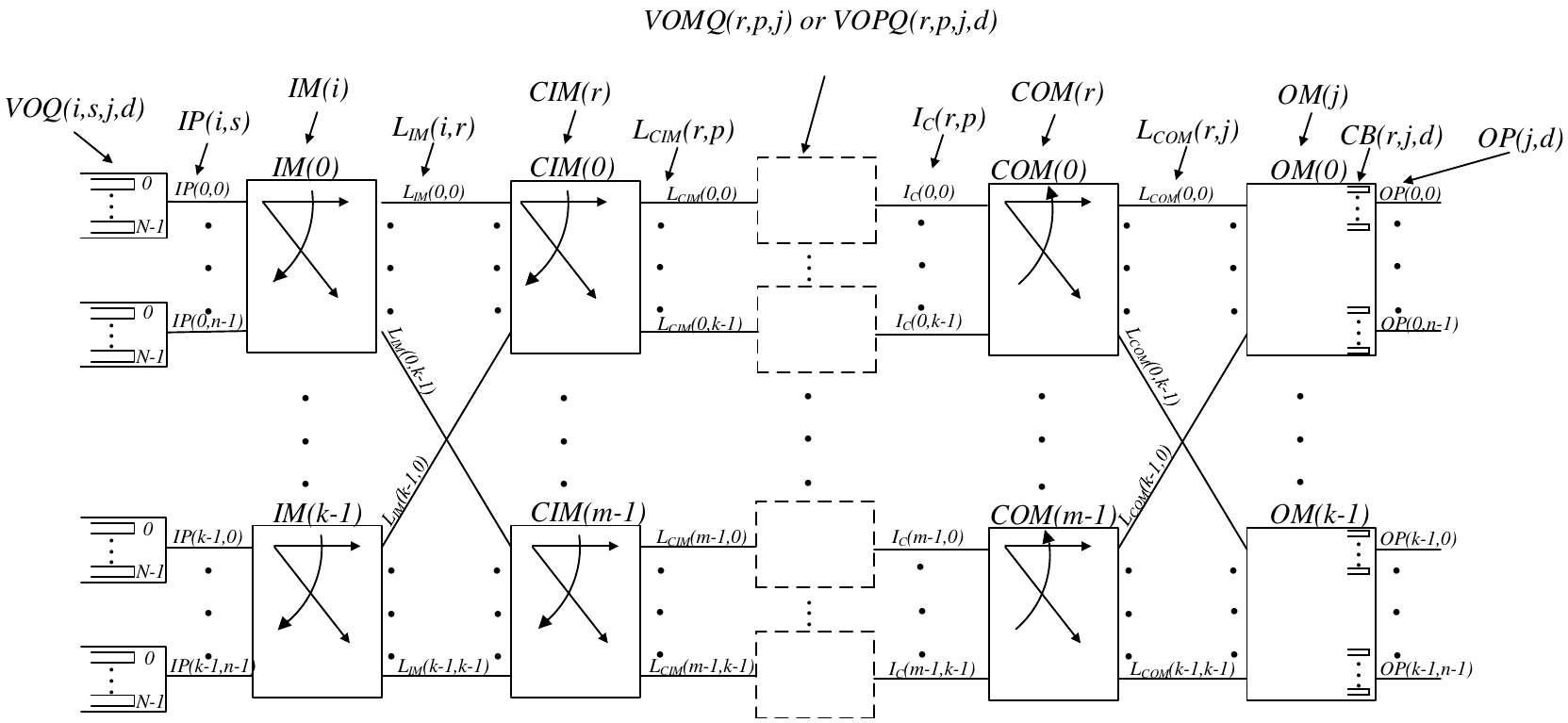}
 	\caption{Split-central buffered load-balancing Clos-network (LBC) switch.}
 	\label{fig:switch}
 \end{figure*}

The LBC switch has $N$ inputs and $N$ outputs, each denoted as $IP(i,s)$ and $OP(j,d)$, respectively, where $0 \leq i,~j \leq k-1$, $0 \leq s,~d \leq n-1$, and $N=nk$. Figure \ref{fig:switch} shows the architecture of the LBC switch. This switch has $k$ $n \times m$ IMs and  $k$ $m \times n$ OMs. Each central module is split into two modules called central-input and -output modules, denoted as CIMs and COMs, respectively. The switch has $m$ CIMs and the same number of COMs. Each CIM and COM is a $k \times k$ switch. In the remainder of this paper, we set $n=k=m$ for symmetry and cost-effectiveness. The IMs, CIMs, and COMs are bufferless crossbars while the OMs are buffered ones.

The use of a split central module on this switch enables preserving staggered symmetry and in-order delivery \cite{hu2008joint} by using a pre-determined configuration in the IMs, CIMs and COMs with a mirror sequence between CIMs and COMs. The staggered symmetry and in-order delivery refers to the fact that at time slot $t$, $IP(i,s)$ connects to $COM(r)$ which connects to $OM(j)$. Then at the next time slot $(t+1)$, $IP(i,s)$ connects to $COM((r+1)\mod m)$, which also connects to $OM(j)$. This property enables the configuration of IMs/CIMs and COMs to be easily represented with a pre-determined compound permutation that repeats every $k$ time slots. This property also ensures that cells experience the same amount of delay for uniform traffic and the incorporation of a simple in-sequence mechanism. 
A switch with queues between IMs and CMs but without a split central module may require more complex load balancing and routing configurations to achieve the same objective.

\begin{figure*}[htbp]
	\centering
	\subfigure[VOPQs.]{
		\includegraphics{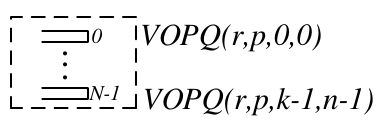}
		\label{fig:switch2}}
	\subfigure[VOMQs]{
		\includegraphics{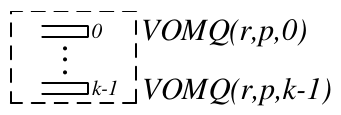}
		\label{fig:switch3}}
	\caption{Split-central buffers with: (a) VOPQs and (b)  VOMQs.}
	\label{fig:switch4}
\end{figure*}

Each input port has $N$ virtual output queues (VOQs), denoted as $VOQ(i,s,j,d)$, to store cells destined to output port $d$ at $OM(j)$.  The combination of IMs and CIMs form a compound stage, called the IM-CIM stage. The COMs and OMs operate as single stages. There are queues placed between CIMs and COMs to store cells coming from an IM and destined to OMs. These central queues may be implemented as virtual output port queues (VOPQs), as shown in Figure \ref{fig:switch2}. Each VOPQ, denoted as $VOPQ(r,p,j,d)$, stores cells coming for $OP(j,d)$ through $L_{CIM}(r,p)$. As an alternative, to reduce the number of VOPQs for a large switch, we consider the use of virtual output module queues (VOMQs) instead, as shown in Figure \ref{fig:switch3}. A VOMQ, denoted as $VOMQ(r,p,j)$, stores cells for all OPs at $OM(j)$. Each of these queues stores cells coming from $L_{CIM}(r, p)$ and destined to $OM(j)$. Compared to VOPQs, VOMQs introduce the possibility of head-of-line (HoL) blocking. However, as we show in Section \ref{proof-no-hol}, such HoL effect is not a concern when the switch is loaded with admissible traffic. The remainder of this paper considers VOMQs, as this option stresses the load-balancing feature of LBC.

Every CIM has $k$ $L_{CIM}$ ports. Every $L_{CIM}(r,p)$ of a CIM is connected to one input $I_C(r,p)$ of the corresponding COM. The LCIM includes a set of $k$ VOMQs, one per OM. Each OP has $m$ crosspoint buffers, each denoted as $CB(r,j,d)$. A flow control mechanism operates between VOMQs and VOQs, and between CBs and VOMQs to avoid buffer overflow and this is described in Section \ref{flow-control-lbc}.
The VOMQs are off-chip. The switch has $N$ LCIMs, and therefore $N$ sets of $k$ VOMQs each. Table \ref{tab:terms} lists the notations used in the description of the LBC switch.

The following is a walk-through description of how the switch operates: After arriving at the IP, a cell is placed at the VOQ corresponding to its destination OP. The IP arbiter selects a VOQ to be served in a round-robin manner. When a VOQ is selected, the HoL cell is forwarded to a VOMQ at the LCIM identified by the current
configuration of the IM and CIM. The VOMQ is the one associated with the OM that includes the destination OP of
the cell. 
When the configuration of the COM permits forwarding to the destination OM, the cell is forwarded to the OM and stored at the crosspoint buffer (CB) allocated for cells from the source COM. The OP arbiter selects CBs based on a round-robin manner. Upon selection of a CB, the HOL cell is forwarded from the CB to the OP.

\begin{table}[htb]
\caption{Notations used in the description of the LBC switch \label{tab:terms}}
\begin{tabular}{l | p{5cm}} \hline \hline
Term & Description \\ \hline \hline
$N$& Number of input/output ports.\\ \hline
$n$ & Number of input/output ports for each IM and OM.\\ \hline
$m$ & Number of CIMs and COMs.\\ \hline
$k$ & Number of IMs and OMs, where $k =\frac{N}{n}$. \\ \hline
$IP(i,s)$ &  Input port $s$ of $IM(i)$, where $0 \leq i \leq k-1, 0 \leq s \leq n-1$.\\ \hline
$IM(i)$ & Input module $i$. \\ \hline
$OM(j)$ & Output module $j$, where $0 \leq j \leq k-1$. \\ \hline
$CIM(r)$ & Central Input Module $r$, where $0 \leq r \leq m-1$.\\ \hline
$COM(r)$ & Central Output Module $r$. \\ \hline
$VOQ(i,s,j,d)$ & VOQ at $IP(i,s)$ that stores cells destined to $OP(j,d)$, where $0 \leq d \leq n-1$.\\ \hline
$L_{IM}(i,r)$ & Output link of $IM(i)$ connected to $CIM(r)$. \\ \hline
$L_{CIM}(r,p)$ & Output port $p$ of $CIM(r)$, where $0 \leq p \leq k-1$. \\ \hline
$I_{C}(r,p)$ & Input port $p$ of $COM(r)$.\\ \hline
$L_{COM}(r, j)$ & Output link of $COM(r)$ connected to $OM(j)$.\\ \hline
$VOMQ(r,p,j)$ & VOMQ at output of CIMs that stores cells destined to $OM(j)$.\\ \hline
$VOPQ(r,p,j,d)$ & VOPQ at output of CIMs that stores cells destined to $OP(j,d)$.\\ \hline
$CB(r,j,d)$ & Crosspoint buffer at $OM(j)$ that stores cells going through $COM(r)$ and destined to $OP(j,d)$. \\ \hline
$OP(j,d)$ & Output port $d$ at $OM(j)$. \\ \hline
\end{tabular}
\end{table}

\subsection{Module Configuration}
\label{sec:operation-of-lbc}

The IMs and CIMs in the IM-CIM stage are configured based on a pre-determined sequence of disjoint permutations, applying one permutation every time slot.  We call a permutation disjoint from the set of permutations if an input-output pair is interconnected in one and only one of the permutations. This pre-determined sequence of permutations repeats every $k$ time slots. Cells at the inputs of IMs are forwarded to the outputs of the CIMs determined by the configuration of that time slot. A cell is then stored in the VOMQ corresponding to its destination OM. 

The COMs follow a configuration similar to that of the CIMs, but in a mirror (i.e., reverse order) sequence. The HoL cell at the VOMQ destined to $OM(j)$ is forwarded to its destination when the input of the COM is connected to the input of the destination $OM(j)$. Else, the HoL cell waits until the required configuration takes place. The forwarded cell is queued at the CB of its destination OP once it arrives in the OM. 
At the OP, a CB (i.e., HoL cell of that queue) is selected from all non-empty CBs by an output arbitration scheme.

The specific configurations of the bufferless modules, IM, CIM, COM, and OM are as follows. 

At time slot $t$,  $IM(i)$ is configured to interconnect input $IP(i,s)$ to $L_{IM}(i,r)$, with:
\begin{equation}
\label{equ1}
r = (s + t) \mod m
\end{equation}

Similarly,  CIM input $L_{IM}(i,r)$ is interconnected to CIM output $L_{CIM}(r,p)$ at time slot $t$ with:
\begin{equation}
\label{equ2}
p = (i + t ) \mod k
\end{equation}

The configuration of COMs is similar to that of IMs, but in a reverse sequence.
At time slot $t$, COM input $I_{C}(r,p)$ is interconnected to output $L_{COM}(r, j)$ with:
\begin{equation}\footnote{$a\; mod\; k = a + (mutiples\; of\; k) > 0$ when $a < 0$ (e.g., -2 mod 5 = 3).}
\label{eq:LII}
j  = (p - t) \mod k
 \end{equation}
Round-robin could also be used to select VOMQs and configure COMs.
OM buffers allow forwarding a cell from a VOMQ to the destination output without requiring port matching \cite{lin2013minimizing}.

\begin{figure}[htbp]
\centering
\subfigure[Time slot 0]{
\includegraphics[width=0.9\columnwidth]{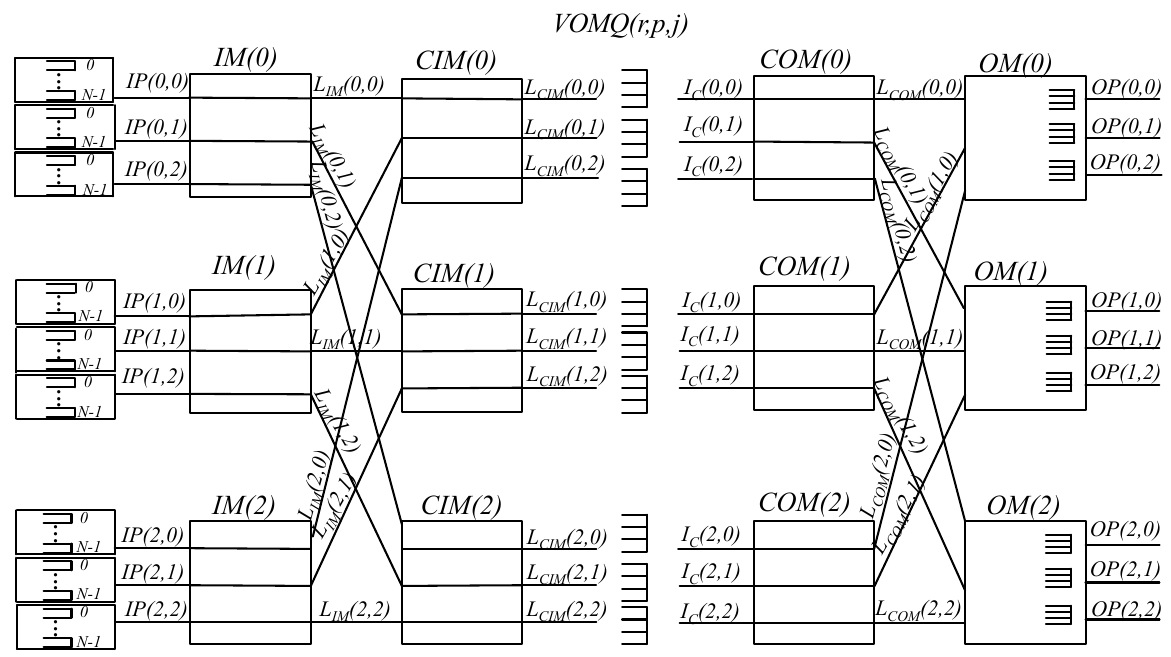}
\label{fig:SwitchConfig1}} 
\subfigure[Time slot 1]{
\includegraphics[width=0.9\columnwidth]{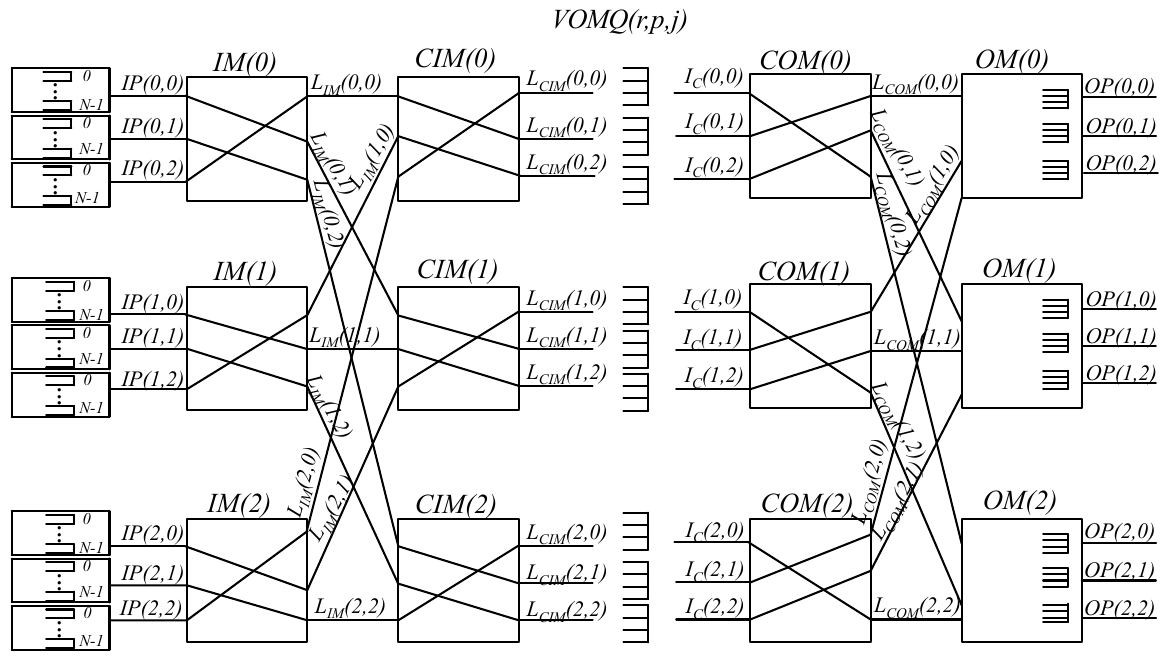}
\label{fig:SwitchConfig2}}
\subfigure[Time slot 2]{
\includegraphics[width=0.9\columnwidth]{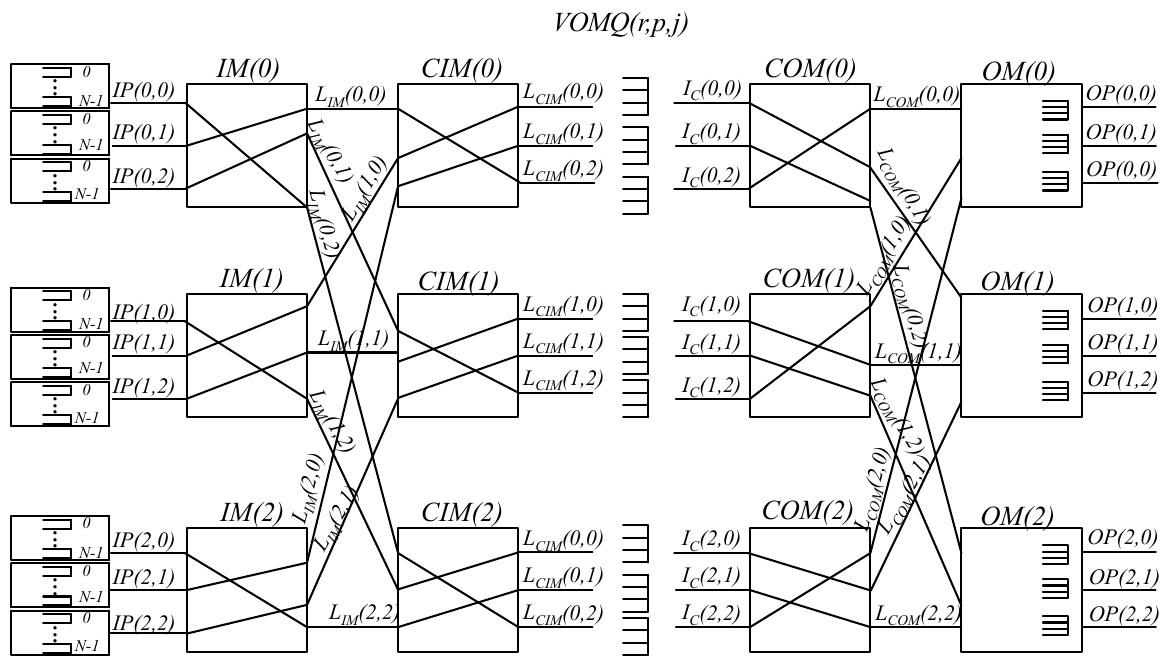}
\label{fig:SwitchConfig3}}
\caption{Configuration example of LBC switch modules.}
\label{fig:switchconfig}
\end{figure}

Figure \ref{fig:switchconfig} shows an example of the configuration of a $9 \times 9$ LBC switch. As $k=3$, the example shows the configuration of three consecutive time slots, after which the configuration pattern repeats. Because similar connections are set for all the IMs and CIMs and a different connection pattern is set for all COMs at each time slot, Table \ref{table:configuration} describes the configuration on the figure for $IM(0)$, $CIM(0)$, and $COM(0)$ at each time slot. In this example, we use $\rightarrow$ to denote an interconnection. 

\begin{table*}[htp]
\centering
\caption{Example of configuration of modules in a 9 $\times$ 9 LBC switch. \label{table:configuration}}
\begin{tabular}{ |c|c|c|c| }
\hline
\multicolumn{4}{ |c| }{Configuration} \\
\hline
Time slot & $IM(0)$ & $CIM(0)$ & $COM(0)$ \\ \hline
\multirow{3}{*}{$t=0$}
 & $IP(0,0) \rightarrow L_{IM}(0,0)$ & $L_{IM}(0,0) \rightarrow L_{CIM}(0,0)$ & $I_c(0,0) \rightarrow L_{COM}(0,0)$\\
  & $IP(0,1) \rightarrow L_{IM}(0,1)$ & $L_{IM}(1,0) \rightarrow L_{CIM}(0,1)$ & $I_c(0,1) \rightarrow L_{COM}(0,1)$\\
  & $IP(0,2) \rightarrow L_{IM}(0,2)$ & $L_{IM}(2,0) \rightarrow L_{CIM}(0,2)$ & $I_c(0,0) \rightarrow L_{COM}(0,2)$\\ \hline
\multirow{3}{*}{$t=1$} 
 & $IP(0,0) \rightarrow L_{IM}(0,1)$ & $L_{IM}(0,0) \rightarrow L_{CIM}(0,1)$ & $I_c(0,0) \rightarrow L_{COM}(0,2)$\\
  & $IP(0,1) \rightarrow L_{IM}(0,2)$ & $L_{IM}(1,0) \rightarrow L_{CIM}(0,2)$ & $I_c(0,1) \rightarrow L_{COM}(0,0)$\\
  & $IP(0,2) \rightarrow L_{IM}(0,0)$ & $L_{IM}(2,0) \rightarrow L_{CIM}(0,0)$ & $I_c(0,2) \rightarrow L_{COM}(0,1)$\\ 
\hline
\multirow{3}{*}{$t=2$} 
 & $IP(0,0) \rightarrow L_{IM}(0,2)$ & $L_{IM}(0,0) \rightarrow L_{CIM}(0,2)$ & $I_c(0,0) \rightarrow L_{COM}(0,1)$\\
  & $IP(0,1) \rightarrow L_{IM}(0,0)$ & $L_{IM}(1,0) \rightarrow L_{CIM}(0,0)$ & $I_c(0,1) \rightarrow L_{COM}(0,2)$\\
  & $IP(0,2) \rightarrow L_{IM}(0,1)$ & $L_{IM}(2,0) \rightarrow L_{CIM}(0,1)$ & $I_c(0,2) \rightarrow L_{COM}(0,0)$\\ 
 \hline
\hline
\end{tabular}
\end{table*}

\subsection{Arbitration at Output Ports}
\label{OP-arbitration_lbc}

An output port arbiter selects a HoL cell from the crosspoint buffers in a round-robin fashion. Because there is one cell from each flow at these buffers, out-of-sequence forwarding is not a concern at this stage. We discuss this case in Section \ref{sec:proof}. Here, a flow is the set
of cells from $IP(i,s)$ destined to $OP(j,d)$. The round-robin schedule ensures fair service for different flows. 

\subsection{In-sequence Cell Forwarding Mechanism}
\label{sec:in-sequence-forwarding} 

The proposed in-sequence forwarding mechanism for the LBC switch is based on 
holding cells of a flow at the VOQs so that no younger cell is forwarded from VOMQs to OPs before any given cell of the same flow. 
The policy used for holding cells at an IP is as follows: No cell of flow $y$ at the IP is forwarded to a VOMQ for $\delta k$ time slots after cell $\tau$ of the same flow has been forwarded to a VOMQ, whose occupancy is $\delta$ cells at the time of arrival  in the VOMQ.    
For a cell that arrives at an empty VOMQ, $\delta=0$.
The flow control mechanism keeps IPs informed about VOMQ occupancy as discussed in Section \ref{flow-control-lbc}.

\begin{figure}[htb]
  \centering
	\includegraphics[width=2.7in]{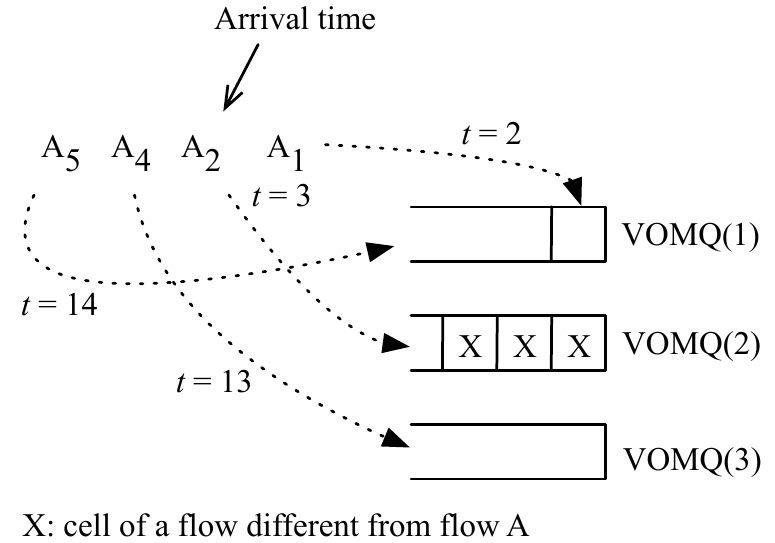}
	\caption{Example of the operation of the proposed in-sequence forwarding mechanism.}
	\label{fig:example-in-order}
\end{figure}

Figure \ref{fig:example-in-order} shows an example of this forwarding mechanism for flow $A$. Cells from flow $A$ are denoted as $A_t$, where $t$ is the cell arrival time. In this example, cells arrive at time slots 1, 2, 4, and 5, and they are denoted as $A_1$, $A_2$, $A_4$, and $A_5$, respectively. VOMQ$(k)$ denotes the $k$th VOMQ to where cells are forwarded. Here, the ``X'' mark indicates that the buffer at VOMQ$(k)$ is occupied by cells from other flows. Assuming $k=3$ and no other cell arrival or departure during this time period, $A_1$ is the first cell of the flow with arrival time $t=1$ and is sent to VOMQ$(1)$ at time slot $t=2$. Because VOMQ$(1)$ has no backlogged cells before $A_1$, there is no waiting time for $A_2$. Therefore, $A_2$ is sent to VOMQ$(2)$ at $t=3$. $A_2$ finds three cells already queued, so no cell from this flow is forwarded in $3*3=9$ time slots, or from time slots $t=4$ to $t=12$. After that, $A_4$ is sent to VOMQ$(3)$ at $t=13$. This cell finds no other cell, so $A_5$ is sent to VOMQ$(1)$ at $t=14$. 

\subsection{Implementation of In-sequence Mechanism}
\label{sec:implementation of in-sequence-forwarding} 
Each IP has an input port counter (IPC) for each VOMQ to which it connects. IPCs keep track of the number of cells at these VOMQs. Each IP also has a hold-down timer for each VOQ. The timer is used by the in-sequence forwarding mechanism. The timer is triggered by the IPC count of the VOMQ where the last cell was forwarded. When a cell is forwarded from a VOQ to VOMQ, and the IPC is updated to $\sigma$, this update sets the hold-down timer for that VOQ for $(\sigma-1) k$ time slots, where $\delta = \sigma - 1$.

\subsection{Flow Control}
\label{flow-control-lbc}
There is a flow control mechanism between VOMQs and IPs and another between CBs and VOMQs that extends to IPs. There are fixed connections between each VOMQ and its $k$ corresponding IPs and between each CB and its corresponding $k$ $I_C$s. Each IP has $mk=N$ occupancy counters, IPCs, one per VOMQ. Each VOMQ updates the corresponding $k$ IPCs about its occupancy. A VOMQ uses two thresholds for flow control; pause ($T_{pv}$) and resume ($T_{rv}$), where $T_{pv} > T_{rv}$, in number of cells. When the occupancy of VOMQ, $|VOMQ|$, is larger than $T_{pv}$, the VOMQ signals the corresponding IPs to pause sending cells to it. When the $|VOMQ| < T_{rv}$, the VOMQ signals the  corresponding IPs to resume sending cells to it. Here, $T_{pv}$ is such that $C_{VOMQ}-T_{pv} \geq D_v$, where $C_{VOMQ}$ is the size of the VOMQ and $D_v$ is the flow-control information delay.

Similar to VOMQs, CBs use two thresholds; pause ($T_{pc}$) and resume ($T_{rc}$), where $T_{pc} > T_{rc}$, and $T_{pc}$ is such that $C_{CB}-T_{pc} \geq D_c$, for a CB size, $C_{CB}$, and flow-control information delay between a CB and corresponding IPs, $D_c$. These CB thresholds work in a similar way as for VOMQs.
Different from IPs, VOMQs have a binary flag to pause/resume forwarding of cells to CBs. 
When the occupancy of a CB, $|CB|$, becomes larger than $T_{pc}$, the CB informs the corresponding VOMQs, and in turn VOMQs inform corresponding IPs to pause forwarding cells to the VOMQ for the congested OP.
With IPs paused for traffic to a CB, traffic already at VOMQs can still be forwarded to CBs as long as $|CB|$ is such that $T_{pc} < |CB| < C_{CB}$. When $|CB| < T_{rc}$, the CB signals the corresponding VOMQs to resume forwarding, and in turn, VOMQs signal source IPs to resume forwarding cells for that destination OP.

\subsection{Avoiding HoL Blocking in LBC with VOMQs}
\label{proof-no-hol}
Concerns of HoL blocking, owning to the aggregation of traffic going to different OPs at the same OM at a VOMQ, may arise. However, one must note that this HoL blocking may occur if and only if a CB gets congested. Here, we argue that the efficient load-balancing mechanism and the use of one CB for each COM at an OP avoids congestion of CBs even in the presence of heavy (but admissible) traffic. We also show that CB occupancy does not build up. Let us consider the input traffic matrix, $\mathbf{R_1}$, with input load, $\lambda_{i,s,j,d}$, which gets load-balanced to CIMs at rate of $\frac{1}{m}$. The aggregate traffic arrival rate at an $L_{CIM}$ from all IMs, $R_{LCIM}$, is:
	\begin{equation}
	\label{CB_analysis_1}
	{R_{LCIM}} =  \frac{1}{m}  \displaystyle\sum_{i=0}^{k} \lambda_{i,s,j,d}
	\end{equation}
	Therefore, the traffic arrival rate to a CB from COMs, $R_{CB}$, is:
	\begin{equation}
	\label{CB_analysis_2}
	R_{CB} =  \frac{1}{mk}  \displaystyle\sum_{}^{k} \displaystyle\sum_{i=0}^{k} \lambda_{i,s,j,d}
	\end{equation}
	To test the growth of CBs, we consider three stressing traffic scenarios:
	a) All IPs in the switch have traffic only for OPs in an OM;
	b) all IPs in an IM forward traffic to all OPs in an OM; and
	c) a single flow, with a large rate, going from an IP to a single OP.
	\\
	Then, for a) the largest arrival rate at IPs while being admissible is:
	\begin{equation}
	\label{lam_1}
	\lambda_{i,s,j,d} = \frac{1}{N}
	\end{equation}
	Substituting (\ref{lam_1}) into (\ref{CB_analysis_2}) and $m=n=k$ yields:
	\begin{equation}
	\label{CB_analysis_3}
	{R_{CB}} =  \frac{1}{k^{2}}  \displaystyle\sum_{}^{k} \displaystyle\sum_{i=0}^{k} \frac{1}{N} = \frac{1}{N} = \frac{1}{k^{2}}
	\end{equation}
	Because round-robin is used as selection policy at an OP, the service rate, ${S_{CB}}$, of a CB would be:
	\[\frac{1}{k} \leq {S_{CB}} \leq 1\]
	Yet, while considering the worst case scenario, or:
	\begin{equation}
	\label{CB_analysis_4}
	{S_{CB}} = \frac{1}{k}
	\end{equation}
	Therefore, CB occupancy does not grow because ${S_{CB} > R_{CB}}$.
\\ \\
	For b), the arrival rate at IPs for admissibility is:
	\begin{equation}
	\label{lam_2}
	\lambda_{i,s,j,d} = \frac{1}{k}
	\end{equation}
	Substituting (\ref{lam_2}) into (\ref{CB_analysis_2}) yields:
	\begin{equation}
	\label{CB_analysis_5}
	{R_{CB}} =  \frac{1}{m}\frac{1}{k}  \displaystyle\sum_{}^{k} \displaystyle\sum_{i=0}^{k} \frac{1}{k} = \frac{1}{k}
	\end{equation}
	The service rate would be the same as in (\ref{CB_analysis_4}). Therefore, the CB would not become congested as ${R_{CB}} = {S_{CB}}$.
\\ \\
	For c), the arrival rate at the IP:
	\begin{equation}
	\label{lam_3}
	\lambda_{i,s,j,d} = 1
	\end{equation}
	The traffic arrival rate to an $L_{CIM}$ is:
	\begin{equation}
	\label{CB_analysis_6}
	{R_{LCIM}} =  \frac{1}{m}  \lambda_{i,s,j,d} =  \frac{1}{m} 
	\end{equation}
	The traffic arrival rate to a CB from COMs is:
	\begin{equation}
	\label{CB_analysis_7}
	{R_{CB}} =  \frac{1}{m}\frac{1}{k}  \displaystyle\sum_{}^{k}  = \frac{1}{m} = \frac{1}{k}
	\end{equation}
	Therefore, the CB would not become congested since ${R_{CB}} \leq {S_{CB}}$ for admissible traffic.

\section{Throughput Analysis}
\label{sec:throughput-analysis}
In this section, we analyze the performance of the proposed LBC switch.
Let us denote the traffic coming to the IM-CIM stage, the COM stage, the OMs, OPs, and the traffic leaving LBC as $\mathbf{R_1}$, $\mathbf{R_2}$, $\mathbf{R_3}$, $\mathbf{R_4}$, and ${R_5}$, respectively. Figure~\ref{fig:switch} shows these analysis points.  Here, $\mathbf{R_1}$, $\mathbf{R_2}$, and $\mathbf{R_3}$ are $N \times N$ matrices, $\mathbf{R_4}$ comprises $N$ $m \times 1$ column vectors, and ${R_5}$ comprises $N$ scalars.

The traffic from input ports to the IM-CIM stage, $\mathbf{R_1}$, is defined as:
\begin{equation}
\label{eq:R1}
\mathbf{R_1}=[\lambda_{u,v}]
\end{equation}

Here, $\lambda_{u, v}$ is the arrival rate of traffic from input $u$ to output $v$, where 

\begin{equation}
\label{equ:u}
u=ik+s
\end{equation}
\begin{equation}
\label{equ:v}
v=jm+d
\end{equation} 
and $0 \leq u, v \leq N-1$. 

In the following analysis, we consider admissible traffic, which is defined as:
\begin{equation}
\label{eq:admissible}
 \sum_{u=0} ^ {N-1} \lambda_{u,v} \leq 1, \  \sum_{v=0} ^ {N-1} \lambda_{u,v} \leq 1
\end{equation}
under i.i.d. traffic conditions.
 
The IM-CIM stage of the LBC switch balances the traffic load coming from the input ports to the VOMQs. Specifically, the permutations used to configure the IMs and CIMs interconnect the traffic from an input to $k$ different CIMs, and then to the VOMQs connected to these CIMs. 

$\mathbf{R_2}$ is the traffic directed towards COMs and it is derived from $\mathbf{R_1}$ and the permutations of IMs and CIMs.
The configuration of the combined IM-CIM stage at time slot $t$ that connects $IP(i,s)$ to $L_{CIM}(r,p)$ are represented as an $N \times N$ permutation matrix, $\mathbf{\Pi}(t)=[\pi_{u,\upsilon}]$, where $r$ and $p$ are determined from (\ref{equ1}) and (\ref{equ2}) and the matrix element:

\begin{eqnarray*}
 \pi_{u,\upsilon} = \begin{dcases*}
        1  & for any $u$, $\upsilon= rk+p $\\
        0 & elsewhere. 
        \end{dcases*}
        \label{eq:4a}
\end{eqnarray*}

The configuration of the compound IM-CIM stage can be represented as a compound permutation matrix, 
$\mathbf{P_{1}}$, which is the sum of the IM-CIM permutations over $k$ time slots as follows,
\begin{equation}
\mathbf{P_1} = \displaystyle\sum_{}^{k} \mathbf{\Pi}(t)
\label{eq:P1}
\end{equation}

Because the configuration is repeated every $k$ time slots, the traffic load from the same input going to each VOMQ is $\frac{1}{k}$ of the traffic load of $\mathbf{R_1}$.
Therefore, a row of $\mathbf{R_2}$ is the sum of the row elements of $\mathbf{R_1}$ at the non zero positions of $\mathbf{P_1}$, normalized by $k$.
This is:
\begin{equation}\label{eq:R2}
\mathbf{R_2} = \frac{1}{k}((\mathbf{R_1}*\mathbb{1}) \circ \mathbf{P_1})
\end{equation}
where $\mathbb{1}$ denotes an $N \times N$ unit matrix and $\circ$ denotes element/position wise multiplication.

There are $k$ non-zero elements in each row of $\mathbf{R_2}$.
Here, $\mathbf{R_2}$ is the aggregate traffic in all the VOMQs destined to all OPs. This matrix can be further decomposed into $k$ $N \times N$ submatrices, $\mathbf{R_2}(j)$, each of which is the aggregate traffic at VOMQs designated for $OM(j)$.
\begin{equation}\label{equ:R2_2}
\mathbf{R_2} = \displaystyle\sum_{j=0}^{j=k-1} \mathbf{R_2}(j)
\end{equation}
where $j$ is obtained from (\ref{equ:v}) $\forall \;d$.
The configuration of the COM stage at time slot $t$ that connects $I_{c}(r,p)$ to $L_{COM(r,j)}$ 
can be represented as an $N \times N$ permutation matrix, 
$\mathbf{\Phi}(t)=[\phi_{u,v}]$, and the matrix element;
\begin{eqnarray}
 \phi_{u,v} = \begin{dcases*}
        1  & for any $u$,~$v= jk+r $\\
        0 & elsewhere. 
        \end{dcases*}
        \label{eq:4}
\end{eqnarray}

Similarly, the switching at the COM stage is represented by a compound permutation matrix $\mathbf{P_2}$, which is the sum of $k$ permutations of the COM stage over $k$ time slots.
Here 
\begin{equation}\label{eq:P2}
\mathbf{P_2} = \displaystyle\sum_{}^{k}\mathbf{\Phi}(t)
\end{equation}

The output traffic of COMs going to different
OMs is described by matrix $\mathbf{R_3}(j)$, which is defined as 
\begin{equation}\label{R3}
\mathbf{R_3}(j) = \mathbf{R_2}(j) \circ \mathbf{P_2}
\end{equation}
where $j$ is obtained from (\ref{equ:v}) $\forall \; d$.
The traffic destined to $OP(j,d)$ at $OM(j)$, $\mathbf{R_3}(j,d)$, is obtained by extracting the traffic elements from $\mathbf{R_3}(j)$, or:
\begin{equation}\label{R3_2}
\mathbf{R_3}(j) =  \displaystyle\sum_{d=0}^{d=k-1} \mathbf{R_{3}}(j,d)
\end{equation}
where $d$ is obtained from (\ref{equ:v}) for the different $j$.

$\mathbf{D_s}$ is an $m \times N$ matrix, built by concatenating $N$ $k \times 1$ vector of all ones, $\vec{1}$, as:
\begin{equation}
\label{eq:ds}
\mathbf{D_s}=[\vec{1}, \cdots, \vec{1}]
\end{equation}
$\vec{A}$ is a $1 \times k$ row vector, built by setting the first element to $1$ and every other element to $0$, or:
\begin{equation}
\label{eq:as}
\vec{A}=[1 \cdots 0]
\end{equation}
$\vec{A_s}$ is an $N \times 1$ column vector, built by concatenating $k$ $\vec{A}$ and taking the transpose, or:
\begin{equation}
\label{eq:as2}
\vec{A_s}=[\vec{A_{s_{1}}}, \cdots, \vec{A_{s_{k}}}]^T
\end{equation}
where $\vec{A_{s_{1}}}=\vec{A_{s_{k}}}=\vec{A}$, such that
\begin{equation}
\label{eq:as3}
\vec{A_s} = [\vec{A}, \cdots, \vec{A}]^T 
\end{equation}
The traffic queued at the CB of an OP, $\mathbf{R_4}(v)$, is the multiplication of $\mathbf{D_s}$, $\mathbf{R_{3}}(j,d)$, and $\vec{A_s}$, or:
\begin{equation}\label{equ9}
\mathbf{R_4}(v) = \mathbf{D_s} * \mathbf{R_{3}}(j,d) * \vec{A_s}
\end{equation}
The traffic leaving an OP, ${R_5}(v)$, is:
\begin{equation}\label{eq:R5_lbc}
{R_5}(v) = (\vec{1})^T * \mathbf{R_4}(v)
\end{equation}
Therefore, $\mathbf{R_5}(v)$ is the sum of the traffic leaving $OP(v)$.

Equations (\ref{eq:R2}), (\ref{equ9}), and (\ref{eq:R5_lbc}) show that the admissibility conditions in (\ref{eq:admissible}) are satisfied by the traffic at the VOMQ, CBs, and OP. Since $\mathbf{R_2}$, $\mathbf{R_4}(v)$, and ${R_5}(v)$  meet the admissibility conditions in (\ref{eq:admissible}), this implies that the sum of the traffic load at each $VOMQ$, $CB$, and $OP$ does not exceed their respective capacities.
From (\ref{equ9}), we can deduce that $\mathbf{R_4}$ is equal to the input traffic $\mathbf{R_1}$, or:
\begin{equation}
\label{theorem1:eq1_lbc}
\mathbf{R_4}(v) = \mathbf{R_1}(v) \; \forall \; v
\end{equation}
From the admissibility of $\mathbf{R_2}$, $\mathbf{R_4}(v)$, ${R_5}(v)$ and (\ref{theorem1:eq1_lbc}), we can infer that the input traffic is successfully forwarded to the output ports.

As discussed in Section \ref{OP-arbitration_lbc}, the output arbiter selects a flow in a round-robin fashion and if no cell of a flow is selected, the OP arbiter moves to the next flow. This implies the queues are work conserving which ensures fairness and that cells forwarded to OPs are successfully forwarded out of OPs. Hence, from (\ref{eq:R5_lbc}), we can infer that ${R_5}(v)$ is equal to $\mathbf{R_4}(v)$, or:
\begin{equation}
\label{theorem1:eq2_lbc}
{R_5}(v) = (\vec{1})^T * \mathbf{R_4}(v) \; \forall \; v
\end{equation} 
From (\ref{theorem1:eq1_lbc}) and (\ref{theorem1:eq2_lbc}), we can conclude that LBC successfully forwards all traffic at IPs out of OPs.

The following example shows the different traffic matrices for a 4$\times$4 ($k= 2$) LBC switch.  Let the input traffic matrix be
\[
 \  \mathbf{R_1} =
\begin{bmatrix}
\setlength{\arraycolsep}{3pt}
\lambda_{0,0}  & \lambda_{0,1} & \lambda_{0,2} & \lambda_{0,3} \\
\lambda_{1,0}  & \lambda_{1,1} & \lambda_{1,2} & \lambda_{1,3} \\
\lambda_{2,0}  & \lambda_{2,1} & \lambda_{2,2} & \lambda_{2,3} \\
\lambda_{3,0}  & \lambda_{3,1} & \lambda_{3,2} & \lambda_{3,3} \\
\end{bmatrix}
\]

First, $\mathbf{R_1}$ is decomposed into $\mathbf{R_2}$ 
at the IM-CIM stage.
From (\ref{eq:P1}), the compound permutation matrix for the IM-CIM stage for this switch is:
\[
\mathbf{P_1} =
\begin{bmatrix}
\setlength{\arraycolsep}{3pt}
1  & 0  & 0 & 1\\
0  & 1  & 1 & 0 \\
0  & 1  & 1 & 0 \\
1  & 0  & 0 & 1
\end{bmatrix}
\]

Using (\ref{eq:R2}), we get:
\[
\Scale[0.95]{
\mathbf{R_2} = \frac{1}{2}
\begin{bmatrix}
\setlength{\arraycolsep}{3pt}
\sum_{i=0}^ {3} \lambda_{0,i}  & 0 & 0 & \sum_{i=0}^ {3} \lambda_{0,i}\\
0  & \sum_{i=0}^ {3} \lambda_{1,i} & \sum_{i=0}^ {3} \lambda_{1,i} & 0 \\
0  & \sum_{i=0}^ {3} \lambda_{2,i} & \sum_{i=0}^ {3} \lambda_{2,i} & 0 \\
\sum_{i=0}^ {3} \lambda_{3,i}  & 0 & 0 & \sum_{i=0}^ {3} \lambda_{3,i} \\
\end{bmatrix}
}
\]

From (\ref{equ:R2_2}), the traffic matrix at VOMQs destined for the different OMs are:\\
\[
\Scale[0.95]{
\mathbf{R_2}(0) = \frac{1}{2}
\begin{bmatrix}
\setlength{\arraycolsep}{3pt}
\lambda_{0,0} + \lambda_{0,1}  & 0 & 0 & \lambda_{0,0} + \lambda_{0,1}\\
0  & \lambda_{1,0} + \lambda_{1,1} & \lambda_{1,0} + \lambda_{1,1} & 0 \\
0  & \lambda_{2,0} + \lambda_{2,1}  & \lambda_{2,0} + \lambda_{2,1}  & 0 \\
\lambda_{3,0} + \lambda_{3,1}   & 0 & 0 & \lambda_{3,0} + \lambda_{3,1}  \\
\end{bmatrix}
}
\]
\[
\Scale[0.95]{
\mathbf{R_2}(1) = \frac{1}{2}
\begin{bmatrix}
\setlength{\arraycolsep}{1pt}
\lambda_{0,2} + \lambda_{0,3}  & 0 & 0 & \lambda_{0,2} + \lambda_{0,3}\\
0  & \lambda_{1,2} + \lambda_{1,3} & \lambda_{1,2} + \lambda_{1,3} & 0 \\
0  & \lambda_{2,2} + \lambda_{2,3}  & \lambda_{2,2} + \lambda_{2,3}  & 0 \\
\lambda_{3,2} + \lambda_{3,3}   & 0 & 0 & \lambda_{3,2} + \lambda_{3,3}  \\
\end{bmatrix}
}
\]

The rows of $\mathbf{R_2}(v)$ represent the traffic from IPs, and the columns represent $VOMQ(r,p,j)$ at $I_C(r,p)$.
From (\ref{eq:P2}), the compound permutation matrix for the COM stage for this switch is:
\[
\mathbf{P_2} =
\begin{bmatrix}
\setlength{\arraycolsep}{3pt}
1  & 0  & 1 & 0\\
1  & 0  & 1 & 0 \\
0  & 1  & 0 & 1 \\
0  & 1  & 0 & 1
\end{bmatrix}
\]
From (\ref{R3}) and (\ref{R3_2}),  the traffic forwarded to an OP is:
\[
\Scale[0.95]{
\mathbf{R_3}(0,0) = \frac{1}{2}
\begin{bmatrix}
\setlength{\arraycolsep}{3pt}
\lambda_{0,0} & 0 & 0 & \lambda_{0,0} \\
0  & \lambda_{1,0}  & \lambda_{1,0} & 0 \\
0  & \lambda_{2,0} & \lambda_{2,0} & 0 \\
\lambda_{3,0} & 0 & 0 & \lambda_{3,0} \\
\end{bmatrix}
}
\]
\[
\Scale[0.95]{
\mathbf{R_3}(0,1) = \frac{1}{2}
\begin{bmatrix}
\setlength{\arraycolsep}{3pt}
\lambda_{0,1}  & 0 & 0 & \lambda_{0,1}\\
0  &  \lambda_{1,1} & \lambda_{1,1} & 0 \\
0  & \lambda_{2,1}  & \lambda_{2,1}  & 0 \\
 \lambda_{3,1}   & 0 & 0 & \lambda_{3,1}  \\
\end{bmatrix}
}
\]
\[
\Scale[0.95]{
\mathbf{R_3}(1,0) = \frac{1}{2}
\begin{bmatrix}
\setlength{\arraycolsep}{3pt}
\lambda_{0,2}  & 0 & 0 & \lambda_{0,2} \\
0  & \lambda_{1,2} & \lambda_{1,2} & 0 \\
0  & \lambda_{2,2} & \lambda_{2,2} & 0 \\
\lambda_{3,2} & 0 & 0 & \lambda_{3,2} \\
\end{bmatrix}
}
\]
\[
\Scale[0.95]{
\mathbf{R_3}(1,1) = \frac{1}{2}
\begin{bmatrix}
\setlength{\arraycolsep}{3pt}
 \lambda_{0,3}  & 0 & 0 & \lambda_{0,3}\\
0  & \lambda_{1,3} & \lambda_{1,3} & 0 \\
0  & \lambda_{2,3}  & \lambda_{2,3}  & 0 \\
 \lambda_{3,3}   & 0 & 0 & \lambda_{3,3}  \\
\end{bmatrix}
}
\]
The rows of $\mathbf{R_3}(j,d)$ represent the traffic from $VOMQ(r,p,j)$ at $I_C(r,p)$ and the columns represent $L_{COM}(r,j)$.
$\mathbf{D_{S}}$ and $\vec{A_s}$ are obtained from (\ref{eq:ds}) and (\ref{eq:as3}), respectively, as:

\[ \mathbf{D_s} =
\begin{bmatrix}
1 & 1 & 1 & 1 \\
1 & 1 & 1 & 1
\end{bmatrix}
\]
\[\vec{A_s} = [1\; 0 \;1 \;0]^T\]
The traffic forwarded from $CB$s to the corresponding $OP$ is obtained from (\ref{equ9}): 
\[
\mathbf{R_4}(0) = \frac{1}{2}
\begin{bmatrix}
\setlength{\arraycolsep}{3pt}
\sum_{i=0}^ {3} \lambda_{i,0}  \\
\sum_{i=0}^ {3} \lambda_{i,0} \\
\end{bmatrix}
\]

\[
\mathbf{R_4}(1) = \frac{1}{2}
\begin{bmatrix}
\setlength{\arraycolsep}{3pt}
\sum_{i=0}^ {3} \lambda_{i,1}  \\
\sum_{i=0}^ {3} \lambda_{i,1} \\
\end{bmatrix}
\]

\[
\mathbf{R_4}(2)= \frac{1}{2}
\begin{bmatrix}
\setlength{\arraycolsep}{3pt}
\sum_{i=0}^ {3} \lambda_{i,2}  \\
\sum_{i=0}^ {3} \lambda_{i,2} \\
\end{bmatrix}
\]

\[
\mathbf{R_4}(3)= \frac{1}{2}
\begin{bmatrix}
\setlength{\arraycolsep}{3pt}
\sum_{i=0}^ {3} \lambda_{i,3}  \\
\sum_{i=0}^ {3} \lambda_{i,3} \\
\end{bmatrix}
\]
The rows of $\mathbf{R_4}(v)$ represent the traffic from $COM(r)$.
Using (\ref{eq:R5_lbc}), we obtain the sum of the traffic leaving the OP, or:\\
\[
{R_5}(0)=
\sum_{i=0}^ {3} \lambda_{i0}
\]
\[
{R_5}(1)=
\sum_{i=0}^ {3} \lambda_{i1}
\]
\[
{R_5}(2)=
\sum_{i=0}^ {3} \lambda_{i2}
\]
\[
{R_5}(3)=
\sum_{i=0}^ {3} \lambda_{i3}
\]

We use the traffic analysis of this section to demonstrate that the LBC switch achieves 100\% throughput under admissible traffic. This demonstration is provided in Appendix \ref{appendix:stability}.

\section{Analysis of In-Sequence Service}
\label{sec:proof}
	
In this section, we demonstrate that the LBC switch forwards cells in sequence through the proposed in-sequence forwarding mechanism.
		
Table \ref{tab: term2} lists the definition of terms used in the discussion of the properties of the proposed LBC switch. 
Here, $c_{y, \tau}(i,s,j,d)$ denotes the $\tau$th cell of traffic flow $y$, which comprises cells going from $IP(i,s)$ to $OP(j,d)$ with arrival time $t_x$. 
In addition, $t_{a_{y,\tau}}$ denotes the arrival time of $c_{y,\tau}$, and $q_{1_{y,\tau}}$, $q_{2_{y,\tau}}$, and $q_{3_{y,\tau}}$ denote the 
queuing delays experienced by $c_{y,\tau}$ at $VOQ(i,s,j,d)$, $VOMQ(r,p,j)$, and 
$CB(r,j,d)$, respectively. The departure times of $c_{y,\tau}$ from these queues are denoted as 
$d_{1_{y,\tau}}$, $d_{2_{y,\tau}}$, and $d_{3_{y,\tau}}$, respectively. 
In this paper, we consider admissible traffic as defined in (\ref{eq:admissible}).

Here, we claim that the LBC switch forward cells in sequence to the output ports, through the following theorem.
\vspace{0.1in}
\begin{theorem}
{\it For any two cells $c_{y, \tau}(i,s,j,d)$ and $c_{y, \tau'}(i,s,j,d)$, where $\tau < \tau'$, $c_{y, \tau}(i,s,j,d)$ departs the destination output port before $c_{y, \tau'}(i,s,j,d)$.}
\end{theorem}

\begin{table}[htb]
\caption{Notations for in-sequence analysis. \label{tab: term2}}
\begin{tabular}{l | p{7cm}} \hline \hline
$c_{y,\tau}$ & The $\tau$th cell of flow $y$ from $IP(i,s)$ to $OP(j,d)$.\\
$t_{a_{y,\tau}}$        & Arrival time of $c_{y,\tau}$ in $VOQ(i,s,j,d)$ at $IP(i,s)$.\\
$q_{1_{y,\tau}}$      & Queuing delay of $c_{y,\tau}$ at $VOQ(i,s,j,d)$.\\
$d_{1_{y,\tau}}$      & Departure time of $c_{y,\tau}$ from $VOQ(i,s,j,d)$ at $IP(i,s)$.\\
$q_{2_{y,\tau}}$      & Queuing delay of $c_{y,\tau}$ at $VOMQ(r,p,j)$.\\
$d_{2_{y,\tau}}$      & Departure time of $c_{y,\tau}$ from $VOMQ(r,p,j)$ at $L_{COM}(r,j)$.\\
$q_{3_{y, \tau}}$      & Queuing delay of $c_{y,\tau}$ at $CB(r,j,d)$ of $OP(j,d)$.\\
$d_{3_{y,\tau}}$      & Departure time of $c_{y,\tau}$ from $CB(r,j,d)$.\\ \hline \hline
\end{tabular}
\end{table}	 

This theorem is sectioned into the following lemmas.
		 
\begin{lemma}
\label{lemma1}
{\it For a single flow traversing the LBC switch, any cell of the flow experiences the same delay. This is, let $t_d$ be the delay experienced by a cell. Then, $t_{d_{y,\tau}}=\gamma ~~ \forall ~ \tau$}, where $\gamma$ is a positive constant.
\end{lemma}

A constant delay for each cell implies that cells depart the switch in the same order they arrived under the conditions of this lemma.

\vspace{0.1in}
\begin{lemma}  
\label{lemma:lemma2}
{\it For any number of flows traversing the LBC switch, cells from the same flow arrive at the OM in sequence.}
\end{lemma}

\vspace{0.1in}
\begin{lemma}
\label{lemma3}
{\it For any number of flows traversing the LBC switch, the cells of each flow arrive and are cleared at the output port (OP) in the same order the cells arrived at the input port (IP).}		
\end{lemma}

Appendix \ref{sec:proof-appendix} presents the proofs of these lemmas.

\section{Performance Analysis}
\label{sec:simulation-performance}

We evaluated the performance of the LBC switch through computer simulation under both uniform and nonuniform 
traffic models. We also compared the performance of the proposed switch with that of an output-queued (OQ) switch, a 
high-performing Memory-Memory-Memory Clos-network (MMM) switch, and an MMM switch with extended 
memory (MM$^e$M). The MMM switch uses forwarding arbitration schemes to select cells from the buffers in 
the previous stage modules and is agnostic to cell sequence, therefore delivering high switching performance. We considered switches with sizes $N = \{64, 256\}$.
 
\subsection{Uniform Traffic}
 
We evaluated the LBC, OQ, MMM, and  MM$^e$M switches under uniform traffic with Bernoulli and bursty arrivals. Figures~\ref{fig:Uniform delay_64} and {\ref{fig:Uniform delay} show the average delay under uniform Bernoulli traffic arrivals for $N=64$ and $N=256$, respectively.
The results in the figures show that the LBC switch achieves 100\% throughput under uniform traffic with Bernoulli arrivals, indicated by the finite and moderate average queuing delay.  
The high throughput performance by the proposed switch is the result of using an efficient load-balancing process 
in the IM-CIM stage. However, this high performance is expected under this traffic pattern as the 
distribution of the incoming traffic is already uniformly distributed.  
 
Figure~\ref{fig:Uniform delay_64} shows that the LBC switch experiences a similar delay as the MM$^e$M switch at high input load.
Figure~\ref{fig:Uniform delay} shows that the LBC switch experiences a slightly higher average delay than the OQ switch. This additional delay in the LBC switch is caused by having cells wait in the VOMQs until a configuration that allows forwarding the cells to their destination output modules takes place. Because MM$^e$M requires an excessive amount of memory to implement the extended set of queues, the measurement of average cell delay cannot be measured for $N$=256 by our simulators. This figure also shows that the LBC switch achieves a lower average delay than the MMM switch with an input load of $0.95$ and larger.

\begin{figure*}[htbp]
   \centering
    \subfigure[Bernoulli uniform traffic, $N$=64]{
   \includegraphics[width=2.25in]{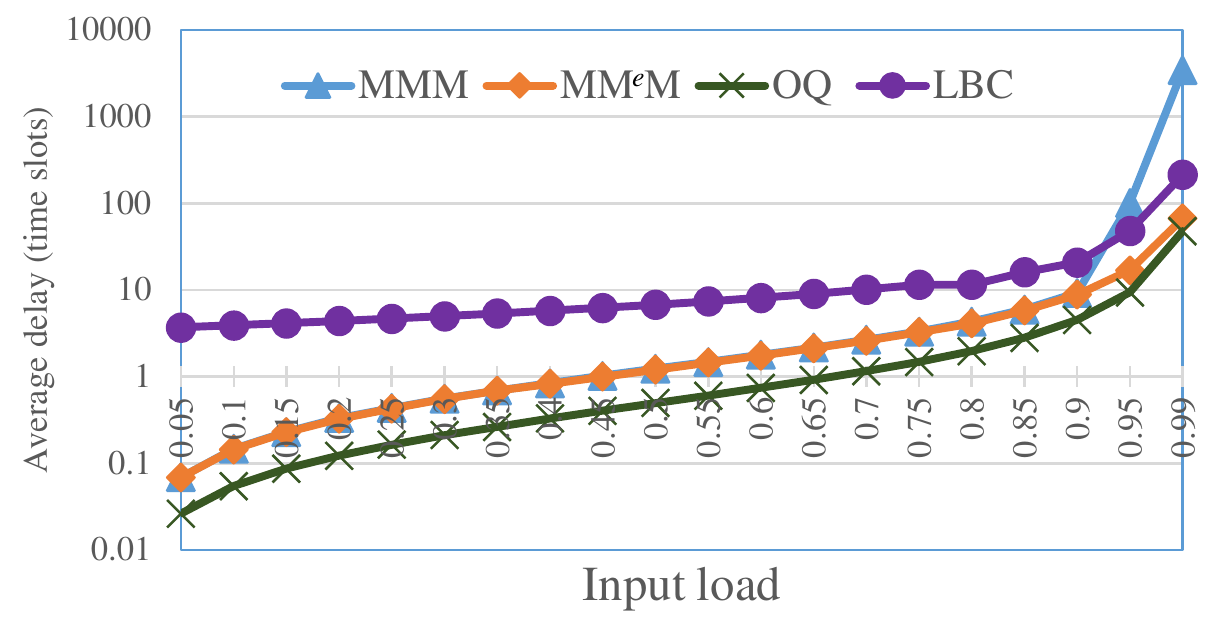}
   \label{fig:Uniform delay_64}}
\subfigure[Bernoulli uniform traffic, $N$=256]{
   \includegraphics[width=2.25in]{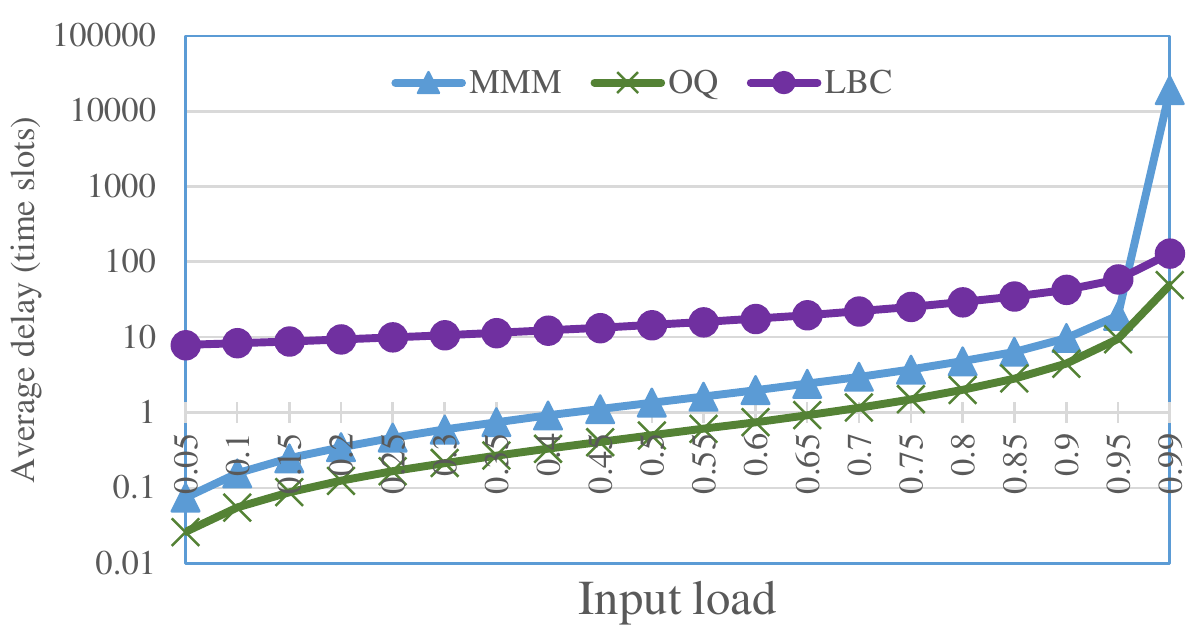}
   \label{fig:Uniform delay}}
   \subfigure[Bursty uniform, $l$=10, $N$=256]{
   \includegraphics[width=2.25in]{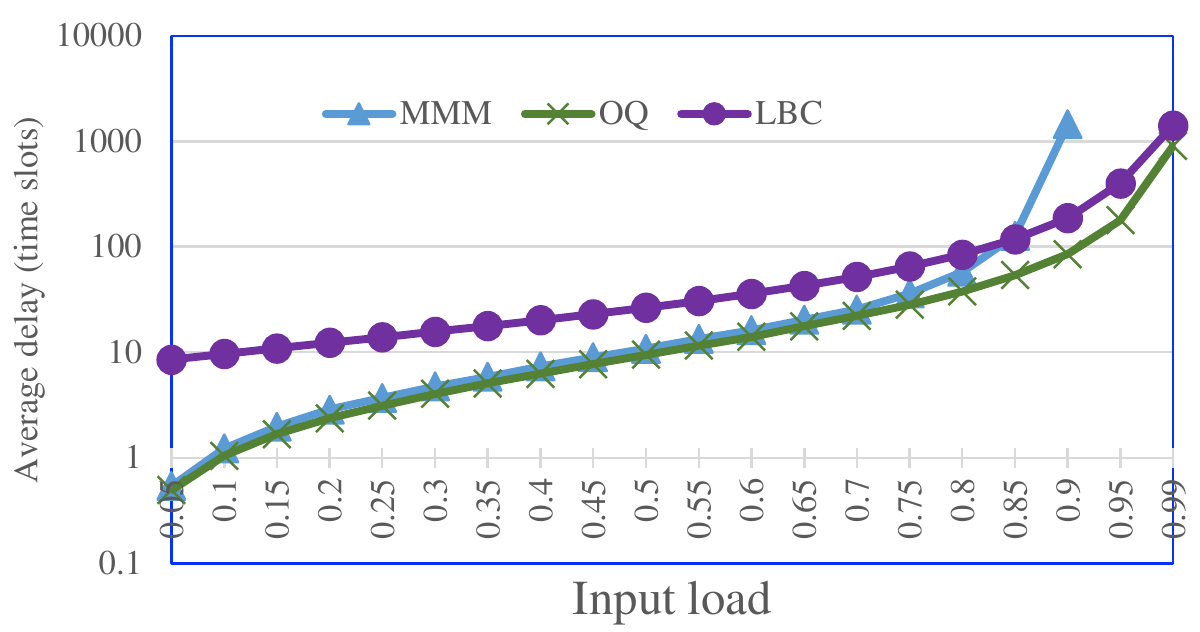}
   \label{fig:bursty10}}
   \subfigure[Bursty uniform, $l$=30]{
   \includegraphics[width=2.25in]{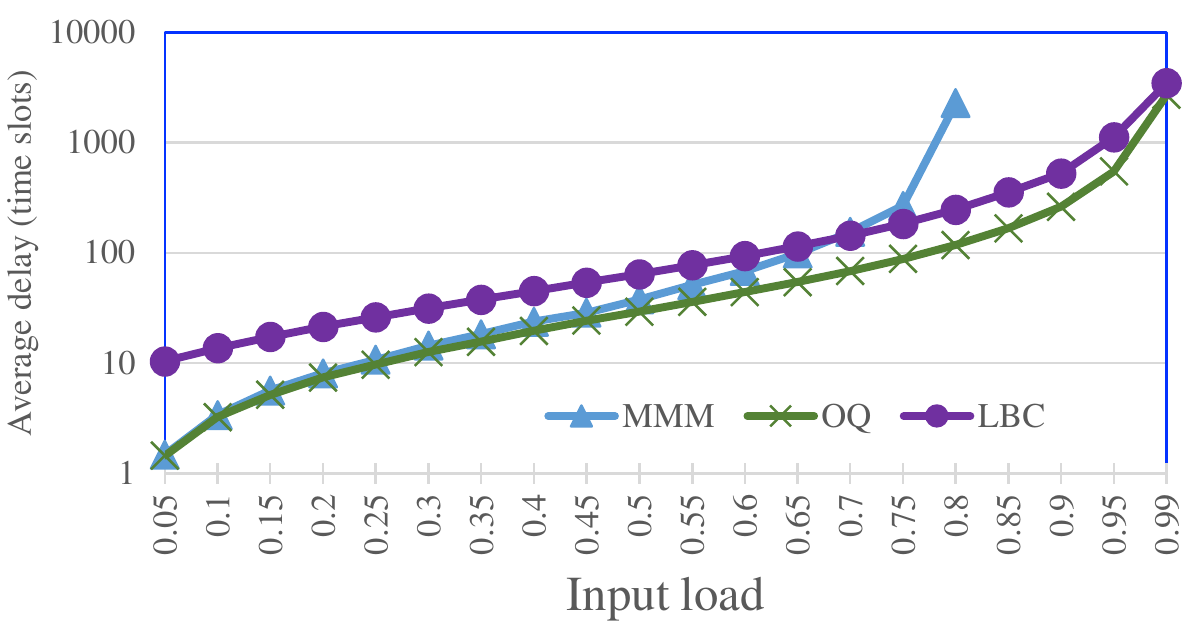}
   \label{fig:bursty30}}
   \subfigure[Unbalanced traffic,$w=0.6$]{
   \includegraphics[width=2.25in]{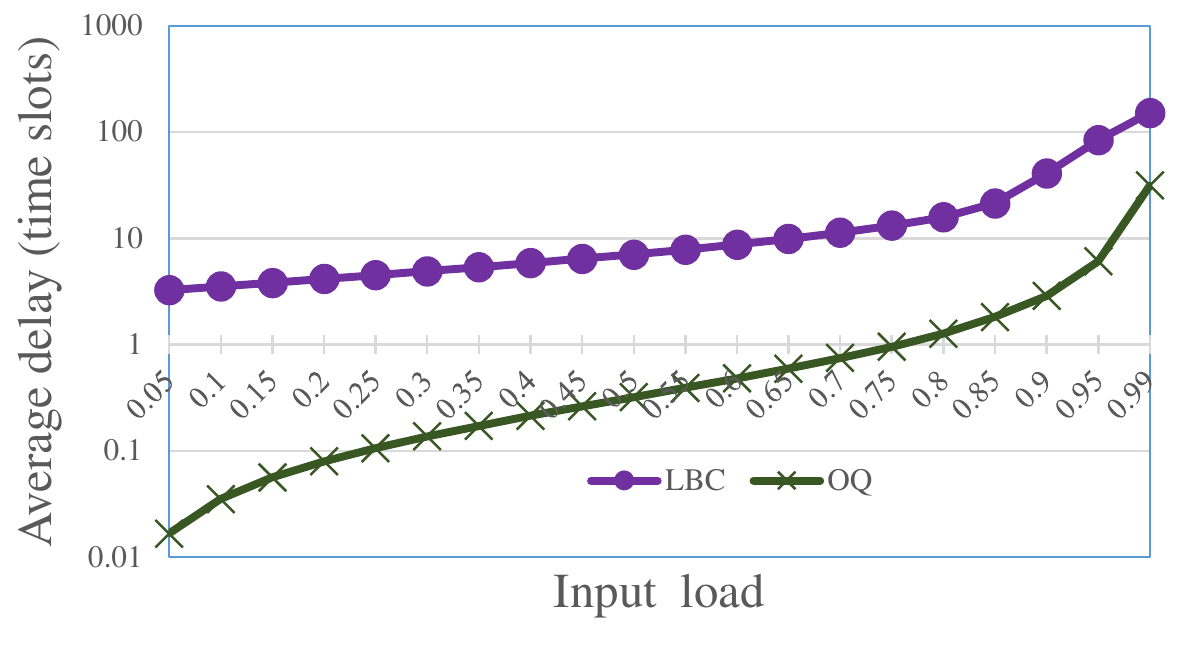}
   \label{fig: Unbalanced Delay}}
   \subfigure[Hot-spot traffic]{
   \includegraphics[width=2.25in]{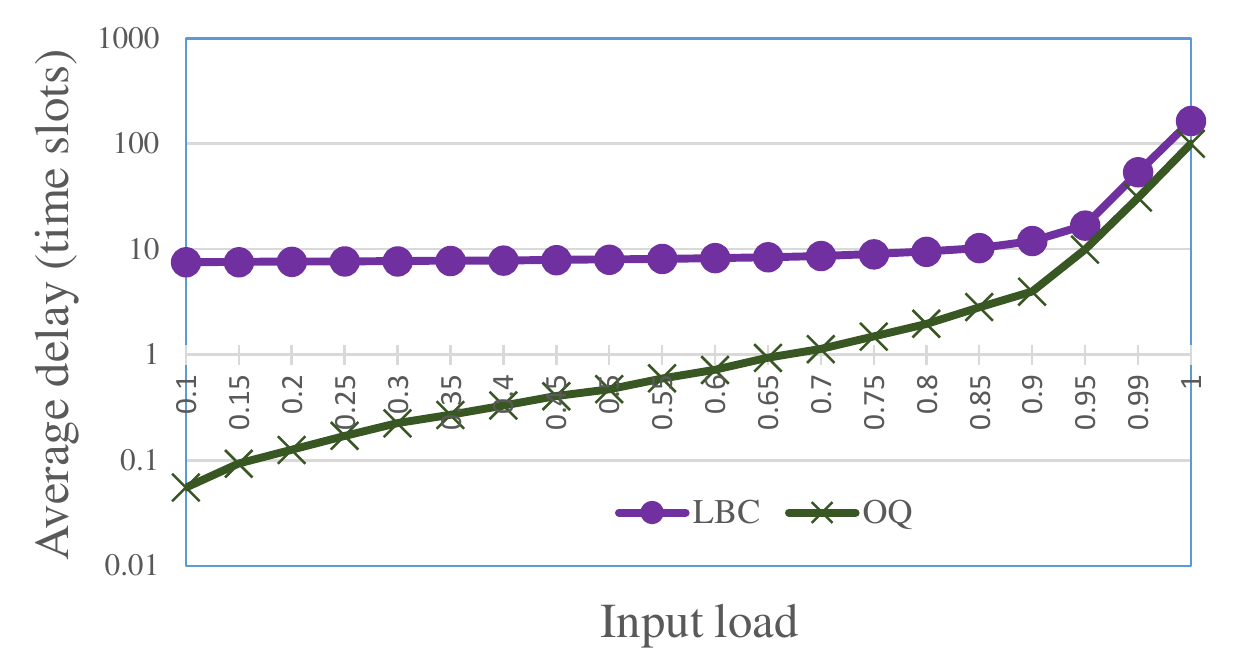}
   \label{fig:Hot-spot delay}}
   \caption{Average queueing delay of LBC switch under uniform traffic: (a) Bernoulli arrivals and $N=64$, (b) Bernoulli arrivals and $N=256$, (c) bursty traffic with average burst  length $l$=10 for $N$=256, and (d) bursty traffic with average burst length $l$=30 for $N$=256, and under nonuniform traffic, such as: (e)  unbalanced traffic with $w=0.6$ for $N$=256 and (f) hot-spot traffic for $N$=256.}
 \label{fig:performance-graphs}
\end{figure*}

 Uniform bursty traffic is modeled as an ON-OFF Markov modulated process, with the average duration of the ON period set as the average burst length, $l$, with $l=\{10, 30\}$ cells. 
 Figures \ref{fig:bursty10} and \ref{fig:bursty30} show the average delay under uniform traffic with bursty arrivals for average burst length of 10 and 30 cells, respectively, for switches with $N$=256.
The results show that the LBC switch achieves 100\% throughput under bursty uniform traffic.  
In contrast, the MMM switch has a throughput of 0.8 and 0.75 for an average burst length of 10 and 30 cells, respectively. Therefore, the LBC switch achieves a performance closer to that of the OQ switch.
There is a very small difference in the delay of the LBC. From this graph, we also observe that the LBC switch achieves 100\% throughput under bursty uniform traffic.
The uniform distributed nature of the traffic and the load-balancing stages help to achieve this high throughput and low queueing delay. The slightly larger average queueing delay of the LBC switch for very small input loads is caused by the predetermined and cyclic configuration of the bufferless modules as some cells wait for a few time slots to be forwarded and this is irrespective of the switch size. Nevertheless, these two figures show that the queueing delay difference between the LBC and the OQ switch is not significant for large input loads.
  
\subsection{Nonuniform traffic}
      
We also compared the performance of the proposed LBC switch with the MMM, MM$^e$M, and OQ switches under unbalanced \cite{cixb,cixob}  and hot-spot patterns as nonuniform traffic.
The unbalanced traffic can be modeled using an unbalanced probability $\omega$ to indicate the load variances for different flows. Consider input port $IP(i,s)$ and output port $OP_(j,d)$ of the LBC switch, the traffic load is determined by
	 \begin{equation}\label{equ25}
	 \rho_{i,s,j,d} = 
	 \begin{dcases}
	 \rho(\omega + {\frac{1 - \omega}{N}}),& \text{if} ~i = j ~\text{and}~s = d,\\
	   \rho{\frac{1 - \omega}{N}},              & \text{otherwise}
	 \end{dcases}
	 \end{equation}    
where $\rho$ is the traffic load for input $IP(i,s)$ and $\omega$ is the unbalanced probability. When $\omega$=0, the input traffic is uniformly distributed and when $\omega$=1, the input traffic is completely directional; traffic from $IP(i,s)$ is destined for $OP(j,d)$.

The simulation results show that the throughput of the LBC switch is 100\% under this traffic pattern for all values of $\omega$, matching those of MMM and MM$^e$M switches, which are also known to achieve high throughput but neglect in-sequence forwarding.
It has been shown that many switches do not achieve high throughput when $w$ is around 0.6 \cite{cixob}. Therefore, we measured the average delay of the LBC switch under this traffic pattern for $\omega$=0.6, as shown in Figure \ref{fig: Unbalanced Delay}, and compared with the OQ switch as this switch is well-known to achieve 100\% throughput. 
As the figure shows, the average delay of the LBC switch is comparable to that of an OQ switch. The load-balancing stage of the LBC switch distributes the traffic uniformly throughout the switch. 

We compared the performance of the proposed LBC switch with the MMM, MM$^e$M, and OQ switches under hot-spot traffic \cite{rojas2016interconnections}.  Hot-spot traffic occurs when all IPs send most or all traffic to one OP. Consider input port $IP(i,s)$ and output port $OP(j,d)$ of the LBC switch, the traffic load is determined by
\begin{equation}\label{hot-spot-lbc}
\rho_{i,s,j,d} = 
\begin{dcases}
\rho(\frac{1}{N}),& \text{for} ~jm+d = h,\\
     0,                               & \text{otherwise}
\end{dcases}
\end{equation}    
where $h$ is the hot-spot OP and $1 \leq h \leq N$.

Our simulation shows that the LBC switch as well as the MMM and MM$^e$M switches achieve 100\% throughput under admissible hot-spot traffic. 

Figure \ref{fig:Hot-spot delay} shows the measured average delay of the LBC switch under this traffic pattern and that of an OQ switch. The figure shows that the average delay of the LBC switch is comparable to that of an OQ switch. This is as a result of effective load-balancing at the IMs, CIMs, and COMs of the multiple flows coming from different inputs.

In addition to the analysis presented in Section \ref{proof-no-hol}, we also simulated the LBC switch under two new traffic patterns, which we believe may stress the occupancy of CBs and therefore increase the likelihood of occurrence of HoL blocking conditions. The traffic patterns are: 
a) $k$ flows from IPs at different IMs, each arriving at a rate of $\frac{1}{k}$ for admissibility, are forwarded to all OPs at one OM. The source IPs of the flows are selected such that they share VOMQs; $i=s$ or $IP(0,0), IP(1,1), \cdots, IP(k-1,n-1) $. 
b) Each IP at an IM forwards cells at rate $\frac{1}{k}$ to each OP at an OM (e.g., $i=j$). Each OP in the destination OM receives traffic from all IPs of one IM. VOMQs are also shared by different flows.
Figures \ref{fig:lbc-compare-all-to-k} and \ref{fig:lbc-compare-mod-to-port} show the average delay under the first and second traffic patterns presented above, respectively. 
The results in the figures show that LBC experiences a finite and moderate average queuing delay, which implies that LBC achieves 100\% throughput under both traffic patterns. We also measured the average CB length and this length does not grow more than one cell, indicating that no CB gets congested. This result is obtained because the load-balancing mechanism spreads a flow to different VOMQs.
	\begin{figure*}[htbp]
		\centering
		\subfigure[$k$ flows from $k$ IMs to all OPs in an OM.]{
			\includegraphics[width=2.25in]{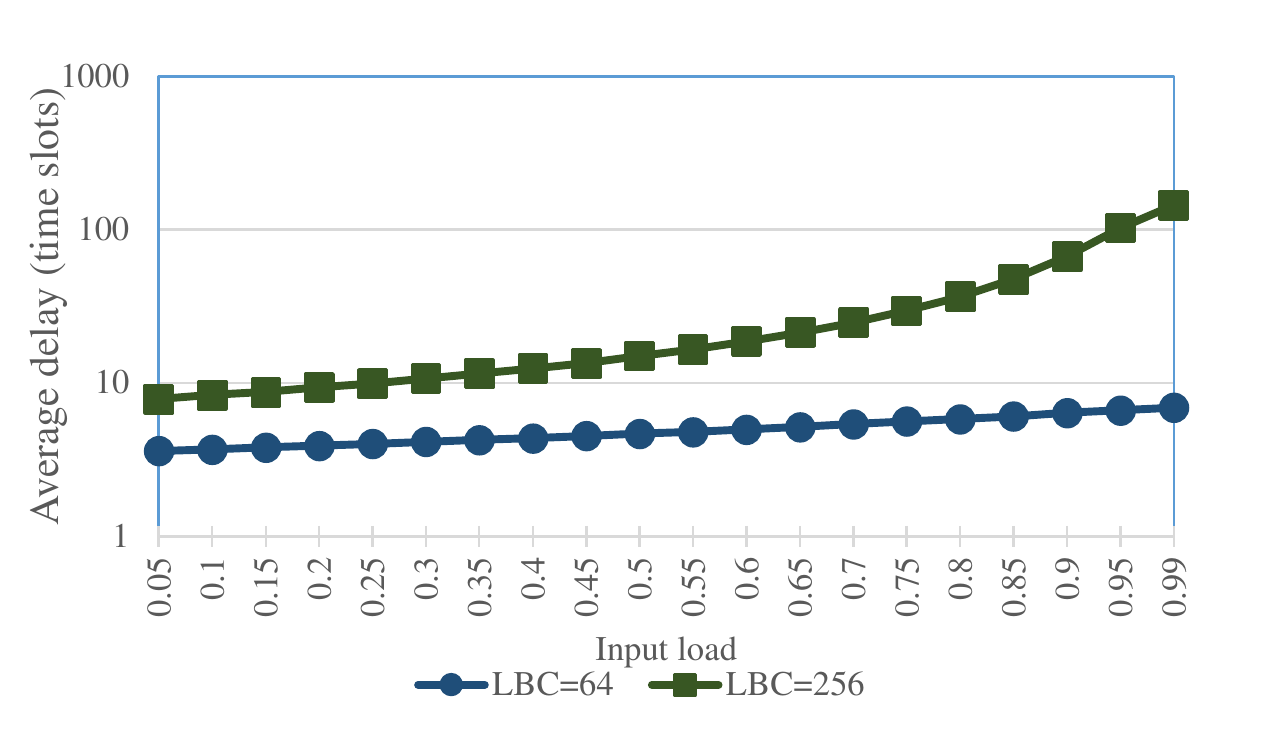}
			\label{fig:lbc-compare-all-to-k}}
		\subfigure[Hot-spot per-module traffic.] {
			\includegraphics[width=2.25in]{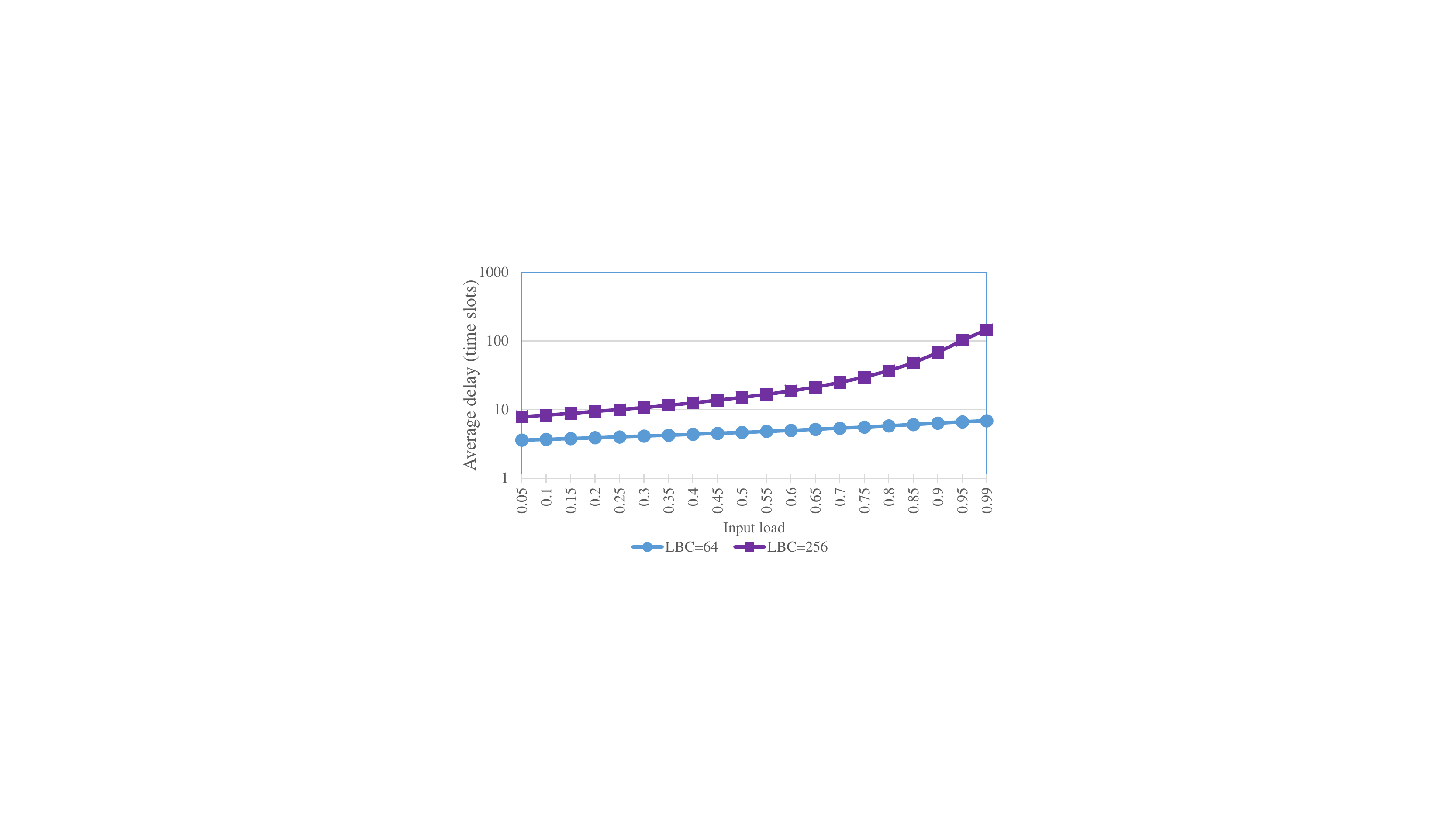}
			\label{fig:lbc-compare-mod-to-port}}
		\caption{Average queueing delay of LBC switch under: (a) $k$ flows from $k$ IMs to all OPs in an OM and (b) Hot-spot per-Module traffic.}
		\label{fig:more-performance-graphs}
	\end{figure*}

\section{Conclusions}
\label{sec:conclusions}
We have introduced a configuration scheme for a split-central-buffered load-balancing Clos-network switch and a mechanism that forwards cells in sequence for this switch. To effectively perform load balancing, the switch has virtual output module queues between these two central stages. With the split central module, the switch comprises four stages, named IM, CIM, COM, and OM. The IM, CIM, and COM stages are bufferless crossbars, while the OMs is a buffered one. All the bufferless modules follow a pre-deterministic configuration while the OM follows a round-robin sequence to forward cells from the CB to the output ports. Therefore, the switch does not have to perform matching in any stage despite having bufferless modules, and the configuration complexity of the switch is minimum, making it comparable to that of MMM switches. We introduce an in-sequence mechanism that operates at the inputs of the LBC switch to avoid out-of-sequence forwarding caused by the central buffers. We modeled and analyzed the operations that each of the stages effects on the incoming traffic to obtain the loads seen by the output ports. We showed that for admissible independent and identically distributed traffic, the switch achieves 100\% throughput. Unlike the existing switching architectures discussed in Section \ref{lbc_intro}, LBC achieves high performance, configuration simplicity, and in-sequence service without memory speedup and central module expansion. In addition, we analyzed the operation of the forwarding mechanism and demonstrated that cells are forwarded in sequence. We showed, through computer simulation, that for all tested traffic, the switch achieved 100\% throughput for uniform and nonuniform traffic distributions.

\appendices
\section{Analysis of In-Sequence Service}
\label{sec:proof-appendix}
	
In this section, we demonstrate the lemmas that support the theorem where we claim that the LBC switch forwards cells in sequence through the proposed in-sequence forwarding mechanism.
				 
{\bf Lemma 1. }{\it For a single flow traversing the LBC switch, any cell of the flow experiences the same delay. This is, let $t_d$ be the delay experienced by a cell. Then, for any cell traversing the LBC switch, $t_{d_{y,\tau}}=\gamma$, where $\gamma$ is a positive constant.}

We analyze first the scenario of a single flow, i.e., $y$, traversing the switch, whose cells arrive back to back, one each time slot. For simplicity but without losing generality, let us also consider empty queues as an initial condition.

{\bf Proof:} 

For any $c_{y,\tau}$, the total delay time is defined as:
\begin{equation}
\label{eq:t_d}
t_{d_{y,\tau}}= q_{1_{y,\tau}}+q_{2_{y,\tau}}+q_{3_{y,\tau}}
\end{equation}
in number of time slots. Here we consider fixed arbitration time at each queue and this delay is included in the queuing delay. 
We are then interested in finding $q_{1_{y,\tau}}$, $q_{2_{y,\tau}}$, and $q_{3_{y,\tau}}$. 

For $q_{1_{y,\tau}}$, under a single-flow scenario, let us consider any two cells of $c_{y, \tau}$ with arrival times $k$ time slots apart, $c_{y, \tau-2k}$ and $c_{y, \tau-k}$, they are forwarded to the same VOMQ. Then, $c_{y, \tau}$ is held at the VOQ (owing to the mechanism to keep cells in sequence at the VOQ) if $c_{y,\tau-k}$ finds one or more cells in the VOMQ, $q_{1_{y,\tau}}$ increases. In this case, the empty queue initial condition makes the waiting factor $\delta=0$. 

On the other hand, an OM is connected to a VOMQ every $k$ time slots as per the configuration scheme of COM. Therefore, 
\begin{equation} 
\label{eq:q2}
q_{2_{y,\tau}} \leq k-1 
\end{equation}

This queuing delay is smaller than the arrival gap between these two cells as:
 
\[  a_{y,\tau-2k}-a_{y,\tau-k}= k~\text{time slots} \]

Therefore, $c_{y, \tau}$ is not backlogged further in VOMQ and there is no impact on the time the cell is held in a VOQ, such that:
\[ q_{1_{y, \tau}}=0 ~~\forall ~y, \tau \]

For $q_{2_{y, \tau}}$, let us now assume that $c_{y, \tau-k}$ arrives at a time that it has to wait $\gamma$ time slots, where $1 \leq \gamma \leq k$, to be forwarded to the destination OM, or 
\[q_{2_{y,\tau-k}}=\gamma  \] 
Then when $c_{y,\tau}$ arrives, $k$ time slots later, it finds exactly the same configuration in the COM as found by $c_{y, \tau-k}$. Because cells arrive consecutively,
\[ q_{2_{y,\tau}}= \gamma ~~\forall ~\tau\]

For $q_{3_{y, \tau}}$, because there is a single flow traversing the switch and the configuration scheme followed by COM, one cell arrives in the CB each time slot and one cell departs OP at the same time slot. Therefore, no cell is backlogged in this case and  
\[  q_{3_{y, \tau}}= 0 \]

From (\ref{eq:t_d}):
\[ t_{d_{y,\tau}} = \gamma ~~\forall ~\tau\]
for empty queues as initial condition. 

It is then easy to see that for any queued cells, $q_{1_{y, \tau}}$ would be increased by $\delta k$ time slots, and $q_{2_{y, \tau}}$ as well as $q_{3_{y, \tau}}$ remain unchanged.

Therefore, all cells of the flow experience the same delay and are forwarded in sequence.

$\blacksquare$	\\

{\bf Lemma 2. } {\it For any number of flows traversing the LBC switch, cells from the same flow arrive at the OM in sequence.}

{\bf Proof:}
Here, we consider the following traffic scenario: There are $k$ flows coming from different IPs, each from a different IM. In each of the flows, cells arrive back to back and are destined to the same OP. Furthermore, the flows have one time slot difference in their arrival times such that the cells with the same sequence number of each different flow are stored in the same VOMQs. Here, each flow consists of $k$ cells. Table \ref{table:arrival-of-k-cells-example} shows an example of the arrival pattern of this traffic scenario for three flows. The table shows the arrival of $k$ cells from $k$ flows at different IPs and IMs that arrive at one time slot apart to enable these flows to be forwarded to the same VOMQ, otherwise the flows would be forwarded to different VOMQs.

\begin{table}[htb]
	\caption{Example of back-to-back arrivals of one burst of $k$ flows.\label{table:arrival-of-k-cells-example}}
	\begin{center}
		\begin{tabular}{ | c | c | c | c |  c | } 
			\hline
			\multicolumn{5}{|c|}{Cell arrival time} \\ \hline
			$t_{x}$ & $t_{x+1}$ & $t_{x+2}$ & $t_{x+3}$ & $t_{x+4}$ \\ 
			\hline
			$c_{1,1}$ & $c_{1,2}$ & $c_{1,3}$ &&  \\ 
			\hline
			& $c_{2,1}$ & $c_{2,2}$ & $c_{2,3}$ & \\ 
			\hline
			&&$c_{3,1}$ & $c_{3,2}$ & $c_{3,3}$  \\
			\hline
		\end{tabular}
	\end{center}
\end{table}
Table \ref{table:arrival-to-middle-queues-k-flows} shows that cells $c_{1,1}$, $c_{1,2}$, $c_{1,3}$, $c_{2,1}$, and $c_{3,1}$ were successfully forwarded to the VOMQ without any blocking.
While the in-sequence mechanism  holds back the cells $c_{2,2}$, $c_{2,3}$, $c_{3,2}$ and $c_{3,3}$ to prevent out-of-sequence, because cells $c_{2,1}$ and $c_{3,1}$ were forwarded to a non-empty VOMQ.

The configuration pattern used in the IMs and CIMs, and the in-sequence mechanism determine the order in which cells arrive to the VOMQs. Table \ref{table:arrival-to-middle-queues-k-flows} shows this order in our example.

\begin{table*}[htb]
	\caption{Time slots in which cells arrive to VOMQs of a single $k$-cell burst. 	\label{table:arrival-to-middle-queues-k-flows}}
	\begin{center}
			\small
			\begin{tabular}{ | c | c | c | c | c | c | c |  c | c | c | c | c |} 
				\hline
				\multicolumn{12}{|c|}{Time slots cells arrive at the VOMQs} \\
				\hline
				$t_{x}$ &$t_{x+1}$ & $t_{x+2}$ & $t_{x+3}$ &$t_{x+4}$ &$t_{x+5}$ &$t_{x+6}$ & $t_{x+7}$ & $t_{x+8}$ & $t_{x+9}$ & $t_{x+10}$ &$t_{x+11}$\\ 
				\hline
				&$c_{1,1}$ & $c_{1,2}$ & $c_{1,3}$ &&&&&&&&\\ 
				\hline
				&& $c_{2,1}$ &&&& $c_{2,2}$ & $c_{2,3}$ &&&&\\ 
				\hline
				&&&$c_{3,1}$ &&&&&&& $c_{3,2}$ &$c_{3,3}$\\
				\hline
			\end{tabular}
	\end{center}
\end{table*}

In such arrival pattern, the departures from VOMQs follow the deterministic configuration of the COMs. Table \ref{table:departures-from-voqcm-example} shows the corresponding departures of the cells from VOMQs of these three flows. 

\begin{table*}[htb]
	\caption{Time slots when cells depart VOMQs in example of the in-sequence forwarding mechanism. \label{table:departures-from-voqcm-example}}
	\begin{center}
			\small
			\begin{tabular}{ | c | c | c | c | c | c | c | c | c | c  | c | c | c |}
				\hline
				\multicolumn{13}{|c|}{Cell departure time slots from VOMQs} \\
				\hline
				$t_{x}$ &$t_{x+1}$ & $t_{x+2}$ &$t_{x+3}$ & $t_{x+4}$ & $t_{x+5}$ &$t_{x+6}$ &$t_{x+7}$ &$t_{x+8}$ & $t_{x+9}$ & $t_{x+10}$ & $t_{x+11}$ & $t_{x+12}$ \\ 
				\hline
				&&&& $c_{1,1}$ & $c_{1,2}$ & $c_{1,3}$&&&&&&\\ 
				\hline
				&&&&&&& $c_{2,1}$ & $c_{22}$ & $c_{2,3}$ &&& \\ 
				\hline
				&&&&&&&&&& $c_{3,1}$ & $c_{3,2}$ & $c_{3,3}$ \\ \hline
			\end{tabular}
	\end{center}
\end{table*}
Table\ref{table:departures-from-voqcm-example} shows that all the cells were forwarded out the VOMQ in the same pattern they arrived and one cell each $k$ time slots because the COM connects to the OM once each $k$ time slots.

Also, let us assume that the first cell of a flow at the $L_{CIM}$ arrives at least one or more time slots before the configuration of the COM allows forwarding the cell to its destination OM. Thus, cells may depart in the following or a few time slot after its arrival. 
A cell then may wait up to $k-1$ time slots for the designated interconnection to take place before being forwarded to the OM. \\

Given $k$ flows, with their $\tau$th cells being $c_{1,\tau}$ to $c_{k,\tau}$, the arrival time of the first arriving cell $c_{1,\tau}$ is:
\begin{equation}\label{t_arrFor1,tau}
t_{a_{1,\tau}} = t_x
\end{equation}
The number of cells at the VOQ, $N_{1}(c_{y,\tau})$, upon the arrival of $c_{1,\tau}$ is:
\begin{equation}\label{N1C1}
N_{1}(c_{1,\tau}) = 0
\end{equation} 
This condition holds because there is no cell at the VOQ when $c_{1,\tau}$ arrives.
Because of (\ref{N1C1}), the queuing delay at the VOQ of $c_{1,\tau}$ is:
\begin{equation}\label{q1tau}
q_{1_{1,\tau}} = 0
\end{equation}
The departure time of a cell  $c_{y,\tau}$ from the VOQ is:
\begin{equation}\label{t_depFor1,tau}
d_{1_{y,\tau}} = t_{a_{y,\tau}} + q_{1_{y,\tau}}
\end{equation}
Using (\ref{t_arrFor1,tau}) to (\ref{t_depFor1,tau}), the departure time of $c_{1,\tau}$ from the VOQ is:
\begin{equation}\label{d1tau}
d_{1_{1,\tau}} = t_x + 1
\end{equation}
Upon arriving at the VOMQ, $c_{1,\tau}$ finds no cell ahead of it. Thus,
the number of cells at the VOMQ, $N_2(c_{1, \tau})$, upon the arrival of $c_{1,\tau}$ is:
\begin{equation}\label{N2C1Tau}
N_{2}(c_{1,\tau})= 0
\end{equation}
Based on the considered traffic pattern, $c_{1,\tau}$ is stored in the VOMQ for additional $k-1$ time slots.
Therefore, 
\begin{equation}\label{q2ForTau}
q_{2_{1,\tau}} = k-1
\end{equation}
The departure time of a cell  $c_{y,\tau}$ from the VOMQ is:
\begin{equation}\label{d2ytau}
d_{2_{y,\tau}} = d_{1_{y,\tau}} + q_{2_{y,\tau}}
\end{equation}\\
Using (\ref{d1tau}), (\ref{q2ForTau}), and (\ref{d2ytau}), the departure time of $c_{1,\tau}$ from the VOMQ is:
\begin{equation}\label{d21tau}
d_{2_{1,\tau}} = t_x + k
\end{equation}

Let us consider now another cell from the same flow, $c_{1,\tau+ \theta}$,
where $0 < \theta < k$, with
\begin{equation}\label{tarrTauplusTheta}
t_{a_{1,\tau + \theta}} = t_x + \theta
\end{equation}
Upon the arrival of $c_{1,\tau+ \theta}$, there is no cell at the VOQ, or:
\begin{equation}\label{N1c1Tau}
N_{1}(c_{1,\tau + \theta}) = 0
\end{equation}
Because of (\ref{N2C1Tau}) and (\ref{N1c1Tau}), the queuing delay at the $VOQ$ for $c_{1,\tau + \theta}$ is:
\begin{equation}\label{q11tautheta}
q_{1_{1,\tau + \theta}} = 0 
\end{equation}
Using (\ref{t_depFor1,tau}), (\ref{tarrTauplusTheta}), and (\ref{q11tautheta}), the departure time of $c_{1,\tau +\theta}$ from the VOQ is:
\begin{equation}\label{dep1TauPlusTheta}
d_{1_{1,\tau + \theta}} = t_x + \theta + 1
\end{equation}
Upon arriving at the VOMQ, $c_{1,\tau +\theta}$ finds no cell ahead of it, or:
\begin{equation}\label{N21TauTheta}
N_{2}(c_{1,\tau + \theta}) = 0
\end{equation}
Because of the considered traffic, $c_{1,\tau + \theta}$ is queued extra $k-1$ time slots at the VOMQ, hence:
\begin{equation}\label{q21tautheta}
q_{2_{1,\tau + \theta}} = k-1
\end{equation}
Using (\ref{d2ytau}), and (\ref{dep1TauPlusTheta})  to (\ref{q21tautheta}),
\begin{equation}\label{d21tautheta}
d_{2_{1,\tau + \theta}} = t_x + k + \theta
\end{equation}
Using (\ref{d21tau}),
therefore, 
\begin{equation}\label{d21tautheta_2}
d_{2_{1,\tau + \theta}} = d_{2_{1,\tau}} + \theta
\end{equation}

In general, for $c_{z,\tau}$, where $1 < z \leq k$, the arrival time is
\begin{equation}\label{taztau}
t_{a_{z,\tau}} = t_x + (z-1)
\end{equation}
and upon the arrival of $c_{z,\tau}$ in the VOQ, there is no cell:
\begin{equation}\label{N1cztau}
N_{1}(c_{z,\tau }) = 0
\end{equation}
With (\ref{N1cztau}),
\begin{equation}\label{q1ztau}
q_{1_{z,\tau}} = 0
\end{equation}
Using (\ref{t_depFor1,tau}), (\ref{taztau}) , and (\ref{q1ztau}),
\begin{equation}\label{d1ztau}
d_{1_{z,\tau}} = t_x + z
\end{equation}
However, upon arriving in the VOMQ, $c_{z,\tau}$ finds $\delta$ cells ahead of it, or:
\begin{equation}\label{N2cztau}
N_{2 }(c_{z,\tau}) = \delta
\end{equation}
\begin{equation}\label{delta}
\delta = z - 1
\end{equation}
where $0 < \delta < k$

\begin{equation}\label{q2ztau}
q_{2_{z,\tau}} =  q_{H_{z,\tau}} + (\delta - 1) k + k 
\end{equation}
$q_{H_{z,\tau}}$ is the delay from the HoL cell in the VOMQ on $c_{z,\tau}$. 
$(\delta - 1)k$ is the delay generated from the other $(\delta - 1)$ cells ahead of $c_{z,\tau}$ in the VOMQ.
The extra $k$ time slots is the delay $c_{z,\tau}$ experiences as it waits for the configuration pattern to repeat after the last cell ahead of it is forwarded to the OM.
where 
\begin{equation}\label{d21tau2}
d_{2_{1,\tau}} = d_{1_{z,\tau}} + q_{H_{z,\tau}}
\end{equation}
Using (\ref{d2ytau}), (\ref{q2ztau}), and (\ref{d21tau2}), the departure time of $c_{z,\tau}$ from the VOMQ is:
\begin{equation}\label{d2ztau22}
d_{2_{z,\tau}} = d_{2_{1,\tau}} + \delta k 
\end{equation}
Using (\ref{d21tau}) and (\ref{delta}), then:
\begin{equation}\label{d2ztau_3}
d_{2_{z,\tau}} = t_x + zk
\end{equation}

Let us now consider any other cell from flow $z$, $c_{z,\tau+ \theta}$,
where $0 < \theta < k$.
The time of arrival of the cell $c_{z,\tau+ \theta}$ is:
\begin{equation}\label{taztautheta}
t_{a_{z,\tau + \theta}} = t_x + (z-1) + \theta
\end{equation}
Upon the arrival of $c_{z,\tau+ \theta}$, there could be zero or more at the VOQ, hence:
\begin{equation}\label{N1cztautheta}
N_{1}(c_{z,\tau + \theta}) =\gamma
\end{equation}
where $\gamma$ is the number of cells at the VOQ upon the arrival of $c_{z,\tau + \theta}$ and $0 \leq \gamma < k$.
Using (\ref{N2cztau}) and (\ref{N1cztautheta}), then:

\begin{eqnarray}
q_{1_{z,\tau + \theta}} = \begin{dcases} 
	 \delta k & for ~ \gamma = 0 \\
	 \delta k + \displaystyle\sum_{\sigma=1}^{\theta - 1} q_{1_{z,\tau + \sigma}} & for ~ \gamma > 0
	    \end{dcases}
 \label{eq:q1_z,tau + theta}
\end{eqnarray}
where \[\displaystyle\sum_{\sigma=1}^{\theta - 1} q_{1_{z,\tau + \sigma}}\] is the delay generated from the $\gamma$ cells ahead of $c_{z,\tau + \theta}$ at the VOQ.
Let
\begin{equation}\label{gamma_q}
\gamma_q = \displaystyle\sum_{\sigma=1}^{\theta - 1} q_{1_{z,\tau + \sigma}}
\end{equation}
Using (\ref{t_depFor1,tau}), (\ref{taztautheta}), (\ref{eq:q1_z,tau + theta}), and (\ref{gamma_q}),
then:
\begin{eqnarray}
	d_{1_{z,\tau + \theta}} = \begin{dcases} 
	t_x + (z-1) + \theta + \delta k   & for ~ \gamma = 0 \\
	t_x + (z-1) + \theta + \delta k + \gamma_q  & for ~ \gamma > 0
\end{dcases}
\label{eq:d2_z,tau_2}
\end{eqnarray}
The queuing delay of $c_{z,\tau + \theta}$ at the VOMQ is equal to (\ref{q2ztau}).
Therefore, using (\ref{d2ytau}), (\ref{q2ztau}), and (\ref{eq:d2_z,tau_2}), the departure time of $c_{z,\tau + \theta}$ from the VOMQ is:
\begin{eqnarray}
	d_{2_{z,\tau + \theta}} = \begin{dcases}
		 d_{2_{1,\tau + \theta}} + \delta k & for ~ \gamma = 0 \\
		 d_{2_{1,\tau + \theta}} + \delta k + \gamma_q & for ~ \gamma >  0
		    \end{dcases}
	 \label{eq:d2_z,tau_3}
 \end{eqnarray}
Using (\ref{d21tautheta_2}) and (\ref{delta}), then:		
\begin{eqnarray}
	d_{2_{z,\tau + \theta}} = \begin{dcases}  
		 d_{2_{1,\tau}} + (z-1)k + \theta & for ~ \gamma =  0 \\
  d_{2_{1,\tau}} + (z-1)k + \theta + \gamma_q & for ~ \gamma >  0
    \end{dcases}
\label{eq:d2_z,tau_4}
\end{eqnarray}
Using (\ref{d21tau}), then:
\begin{eqnarray}
	d_{2_{z,\tau + \theta}} = \begin{dcases} 
		 t_x + zk + \theta & for ~ \gamma =  0 \\
         t_x + zk + \theta + \gamma_q & for ~ \gamma >  0
          \end{dcases}
     \label{eq:d2_z,tau_5}
 \end{eqnarray}
From (\ref{d21tautheta_2}),
\begin{equation}\label{d21tautheta_3}
d_{2_{1,\tau + \theta}} - d_{2_{1,\tau}} = \theta
\end{equation}
Using (\ref{d2ztau_3}), gives:
\begin{eqnarray}
d_{2_{z,\tau + \theta}} - d_{2_{z,\tau}} = \begin{dcases} 
\theta & for ~ \gamma =  0 \\
 \theta +\gamma_q & for ~ \gamma >  0
 	\end{dcases}
 \label{eq:d2_z,tau_7}
 \end{eqnarray}

The difference between the departure times of any two cells of a flow from VOMQ is a function of $\theta$, which is the arrival time difference of the two cells. Therefore, cells of a flow are forwarded to the OM in the same order they arrived.

$\blacksquare$	\\
				
{\bf Lemma 3. } {\it For any number of flows traversing the LBC switch, the cells of each flow arrive and are cleared at the output port (OP) in the same order the cells arrived at the input port (IP).}		
\\
\\
In our discussion of this lemma, let us consider the following traffic scenario: The switch has cells from only two flows, each arriving in a different IM (and therefore IP) and both of them are destined to the same OP. In each flow, cells arrive back-to-back, one at each time slot, and the first cell of both flows arrive at a time slot such that the configuration pattern of IM-CIM stage would not enable forwarding them to the COM immediately. With this condition, we analyze how these two flows are kept from affecting each other, and therefore, the sequence in which cells may depart the OP. This traffic scenario may present the greatest opportunity of experiencing out-of-sequence forwarding by any two cells of a flow as cells from these two flows interact at the CBs of the destination OP. Let us also consider empty queues as an initial condition. 

Given flows $y$ and $z$, where the first cells of $y$  and $z$, $c_{y,\tau}$ and  $c_{z,\tau}$, respectively, arrive at their respective VOQs at time slot $t_x$ and the $\theta$th cells, $c_{y,\tau+\theta}$ and  $c_{z,\tau + \theta}$ $\forall$ $\theta$ $\ge$ 1, arrive at time slot $t_x + \theta$. Therefore, according to this lemma $c_{y,\tau}$ and  $c_{z,\tau}$ must be forwarded and cleared from the output port $OP(j,d)$ before $c_{y,\tau+\theta}$ and  $c_{z,\tau+\theta}$, respectively.

{\bf Proof:} 

We analyze the departure time of the cells $c_{y,\tau}$ and  $c_{z,\tau}$ from the CBs.
The arrival times for cells $c_{y,\tau}$ and  $c_{z,\tau}$ is:
\begin{equation}\label{tayztau}
t_{a_{y,\tau}} = t_{a_{z,\tau}} = t_x
\end{equation} 
Upon arriving in the VOQ, $c_{y,\tau}$ and  $c_{z,\tau}$ are placed as HoL cells. Because there are no backlogged cells, hence:
\begin{equation}\label{N1cytaulem3}
N_{1}({c_{y,\tau}}) = 0
\end{equation} 
and
\begin{equation}\label{N1cztaulem3}
N_{1}({c_{z,\tau}}) = 0
\end{equation}
Using (\ref{N1cytaulem3}) and (\ref{N1cztaulem3}), the queuing delays of  $c_{y,\tau}$ and  $c_{z,\tau}$ at the VOQ are:			
\begin{equation}\label{q1cytaulem3}
q_{1}c_{y,\tau} = 0 
\end{equation}
and
\begin{equation}\label{q1cztaulem3}
q_{1}c_{z,\tau} = 0
\end{equation}
Using (\ref{t_depFor1,tau}), (\ref{tayztau}), and (\ref{q1cytaulem3}) the departure time for $c_{y,\tau}$ from the VOQ is:
\begin{equation}\label{d1ytaulem3}
d_{1{y,\tau}} = t_x + 1
\end{equation}
Using (\ref{t_depFor1,tau}), (\ref{tayztau}), and (\ref{q1cztaulem3}) the departure time for $c_{z,\tau}$ from the VOQ is:
\begin{equation}\label{d1ztau_lem3}
d_{1{z,\tau}} = t_x + 1
\end{equation}
Thus, $c_{y,\tau}$ and  $c_{z,\tau}$ are forwarded to the same CIM (so that these two cells would share the same CB) and stored in their respective VOMQ. Because the VOMQs are empty at the time the two cells arrive, hence:
\begin{equation}\label{N2ytaulem3}
N_{2}({c_{y,\tau}}) = 0 
\end{equation}
 and 
 \begin{equation}\label{N2ztaulem3}
 N_{2}({c_{z,\tau}}) = 0
 \end{equation} 
Based on the adopted traffic scenario, $c_{y,\tau}$ and  $c_{z,\tau}$ are held at the VOMQ for $\beta_1$ and $\beta_2$ time slots, respectively, before the configuration pattern enables forwarding them to their destination OM. Here, $1 \leq \beta_1 < k$ and $ 1 \leq \beta_2 < k$. Hence, the queuing delay of $c_{y,\tau}$ at the VOMQ is:
 \begin{equation}\label{q2ytaulem3}
 q_{2_{y,\tau}} = \beta_1
 \end{equation}
 The queuing delay of $c_{z,\tau}$ at the VOMQ is:
 \begin{equation}\label{q2ztaulem3}
 q_{2_{z,\tau}} = \beta_2
 \end{equation}
Assuming $\beta_1 < \beta_2$, hence $c_{y,\tau}$ would be forwarded to the destination OM before $c_{z,\tau}$. From (\ref{d2ytau}), (\ref{d1ytaulem3}), and (\ref{q2ytaulem3}), the departure time of $c_{y,\tau}$ from the VOMQs is:
\begin{equation}\label{d2ytau_lem3}
d_{2{y,\tau}} = t_x + 1 + \beta_1
\end{equation}
From (\ref{d2ytau}), (\ref{d1ztau_lem3}), and (\ref{q2ztaulem3}), the departure time of $c_{z,\tau}$ from the VOMQs is:
\begin{equation}\label{d2ztau_lem3}
d_{2{z,\tau}} = t_x + 1 + \beta_2
\end{equation}

When $c_{y,\tau}$ and  $c_{z,\tau}$ arrive at the OM, they are stored at CBs before being forwarded to the output port.

Let us now consider $c_{y,\tau+1}$ and  $c_{z,\tau+1}$, which arrive at time slot $t_x$ + 1, hence:
\begin{equation}\label{tayztau+1}
t_{a_{y,\tau+1}} = t_{a_{z,\tau+1}} = t_x + 1
\end{equation}
Because there are no cells at the VOQ upon the arrival of $c_{y,\tau+1}$ and $c_{z,\tau+1}$, then:
\begin{equation}\label{N1ytau+1lem3}
N_{1}c_{y,\tau+1} = 0
\end{equation}
 and  
 \begin{equation}\label{N1ztau+1lem3}
 N_{1}c_{z,\tau+1} = 0
 \end{equation}
 With  (\ref{N2ytaulem3}) and (\ref{N1ytau+1lem3}), the queuing delay of  $c_{y,\tau+1}$ at the VOQ is:			 
 \begin{equation}\label{q1ytau+1lem3}
 q_{1{y,\tau+1}} = 0
 \end{equation}
 With  (\ref{N2ztaulem3}) and (\ref{N1ztau+1lem3}), the queuing delay of  $c_{z,\tau+1}$ at the VOQ is:
 \begin{equation}\label{q1ztau+1lem3}
 q_{1{z,\tau+1}} = 0
 \end{equation}
Using (\ref{t_depFor1,tau}), (\ref{tayztau+1}), and (\ref{q1ytau+1lem3}), the departure time of $c_{y,\tau+1}$ from the VOQ is:			 
\begin{equation}\label{d1ytau+1lem3}
d_{1{y,\tau+1}} = t_x + 2
\end{equation}
Using (\ref{t_depFor1,tau}), (\ref{tayztau+1}), and (\ref{q1ztau+1lem3}), the departure time of $c_{z,\tau+1}$ from the VOQ is:	
\begin{equation}\label{d1ztau+1lem3}
d_{1{z,\tau+1}} = t_x + 2
\end{equation}
$c_{y,\tau+1}$ and  $c_{z,\tau+1}$ are forwarded to the same CIM and stored in their respective VOMQs. Based on the traffic scenario $c_{y,\tau+1}$ and  $c_{z,\tau+1}$ are also stored for $\beta_1$ and $\beta_2$ time slots, respectively, at the VOMQs before the configuration pattern of the COM enables forwarding them to the destination OM. 
Hence, the queuing delay of $c_{y,\tau+1}$ and $c_{z,\tau+1}$ at the VOMQ are equal to (\ref{q2ytaulem3}) and (\ref{q2ztaulem3}), respectively.
From (\ref{d2ytau}), (\ref{q2ytaulem3}), and (\ref{d1ytau+1lem3}), the departure time of $c_{y,\tau+1}$  from the VOMQ is:
\begin{equation}\label{d2ytau1lem3}
d_{2{y,\tau+1}} = t_x + 2 + \beta_1
\end{equation}
From (\ref{d2ytau}), (\ref{q2ztaulem3}), and (\ref{d1ztau+1lem3}), the departure time of $c_{z,\tau+1}$  from the VOMQ is:
\begin{equation}\label{d2ztau+1lem3}
d_{2{z,\tau+1}} = t_x + 2 + \beta_2
\end{equation}

Next, we analyze the departure time of the cells from the output port. Because $d_{2{y,\tau+1}} > d_{2{y,\tau}}$ and $d_{2{z,\tau+1}} > d_{2{z,\tau}}$, this means that $c_{y,\tau}$ and  $c_{z,\tau}$ arrive at the output module before $c_{y,\tau+1}$ and  $c_{y,\tau+1}$, respectively.
With the CB initially empty based on the initial condition, then:
\begin{equation}\label{N3ytau}
N_{3}c_{y,\tau} = 0
\end{equation}
With $d_{2{z,\tau}} > d_{2{y,\tau}}$, hence:
\begin{equation}\label{N3ztau}
N_{3}c_{z,\tau} = 0
\end{equation}
With (\ref{N3ytau}) and (\ref{N3ztau}), the queuing delays of $c_{y,\tau}$ and  $c_{z,\tau}$ at the CB are:
\begin{equation}\label{q3ytau}
q_{3{y,\tau}} = 0
\end{equation}
and
\begin{equation}\label{q3ztau}
q_{3{z,\tau}} = 0
\end{equation}
The queuing delay of $c_{y,\tau+1}$ and $c_{z,\tau+1}$ at the CB are equal to (\ref{q3ytau}) and (\ref{q3ztau}).
The departure time of a cell  $c_{c,\tau}$ from the CB is:
\begin{equation}\label{d3ctau}
d_{3_{c,\tau}} = d_{3_{c,\tau}} + q_{3_{c,\tau}}
\end{equation}
Therefore, using (\ref{d2ytau_lem3}), (\ref{q3ytau}), and (\ref{d3ctau}), the departure time of $c_{y,\tau}$ from the output port is:
\[d_{3{y,\tau}} = t_x + 2 + \beta_1\]
Using (\ref{d2ytau1lem3}), (\ref{q3ytau}), and (\ref{d3ctau}), the departure time of $c_{y,\tau+1}$ from the output port is:
\[d_{3{y,\tau+1}} = t_x + 3 + \beta_1\]
Using (\ref{d2ztau_lem3}), (\ref{q3ztau}), and (\ref{d3ctau}), the departure time of $c_{z,\tau}$ from the output port is:
\[d_{3{z,\tau}} = t_x + 2 + \beta_2\]
Using (\ref{d2ztau+1lem3}), (\ref{q3ztau}), and (\ref{d3ctau}), the departure time of $c_{z,\tau+1}$ from the output port is:
\[d_{3{z,\tau+1}} = t_x + 3 + \beta_2\]

Therefore, with $d_{3{y,\tau+1}} > d_{3{y,\tau}}$ and $d_{3{z,\tau+1}} > d_{3{z,\tau}}$, $c_{y,\tau}$ and $c_{z,\tau}$ would depart the output port before $c_{y,\tau+1}$ and $c_{z,\tau+1}$, respectively.  Note that for $N_{1}(c_{y,\tau})>0$, $\delta>0$, such that the cells from the same flow are forwarded with larger time separation from each other, and there are fewer chances that they will be at the CBs at the same time slot. Therefore, this property, as described by this lemma, applies to any two cells of a flow. 

$\blacksquare$	\\

This completes the proof of Theorem 1.

$\blacksquare$	\\

\section{100\% Throughput}
\label{appendix:stability}
In this section we prove that LBC  achieves 100\% throughput by using the analysis presented on Section III. A and the concept of queue stability.  A switch is defined as stable for a traffic pattern if the queue length is bounded and a switch achieves 100\% throughput if it is stable for admissible i.i.d. traffic \cite{mekkittikul1998practical}. With this, we set the following theorem:
\begin{theorem}
	\label{theorem:throughput_lbc}
	{\it LBC achieves 100\% throughput under admissible i.i.d traffic.}
\end{theorem}

\noindent {\bf Proof:}
Here, we consider the queue to be weakly stable if the drift of the queue occupancy from the initial state is a finite integer $\epsilon$ ~ $\forall ~ t$ as $\lim_{t\to\infty}$.
Using the definition above, we show that the queue length of VOQs, VOMQs, and CBs are weakly stable under i.i.d. traffic, and hence, achieves 100\% throughput under that traffic pattern. 

Let us represent the queue occupancy of VOQs at time slot $t$, $\mathbf{N_1}(t)$ as:
\begin{equation}
\label{N_1(n)-1_lbc}
\mathbf{N_1}(t) = \mathbf{N_1}(t-1) + \mathbf{A_1}(t) - \mathbf{D_1}(t)
\end{equation}
where $\mathbf{A_1}(t)$ is the packet arrival matrix at time slot $t$ to VOQs and $\mathbf{D_1}(t)$ is the service rate matrix of VOQs at time slot $t$.  
Solving (\ref{N_1(n)-1_lbc}) with an initial condition $\mathbf{N_1}(0)$, recursively yields:
\begin{equation}
\label{N_1(n)-2_lbc}
\mathbf{N_1}(t) = \mathbf{N_1}(0) + \displaystyle\sum_{\gamma=0}^{t} \mathbf{A_1}(\gamma) - \displaystyle\sum_{\gamma=0}^{t} D_1(\gamma)
\end{equation}
Let us consider $s_{1_{u,v}}(t)$ as the service rate received by the VOQ at $IP(u)$ for $OP(v)$ at time slot $t$ or:
\begin{eqnarray}
 \begin{dcases*}
\frac{1}{N}  \leq s_{1_{u,v}}(t) \leq 1 & for $\delta = 0$ \\
 \frac{1}{\delta Nk} \leq s_{1_{u,v}}(t) \leq \frac{1}{\delta k}  & for $\sigma > 1 $
\end{dcases*}
\label{Dij_lbc}
\end{eqnarray}
Another way to express $D_1(t)$ is:
\begin{equation}\label{eq:D1_lbc}
\mathbf{D_1}(t)=[s_{1_{u,v}}(t)]
\end{equation}
and recalling $\mathbf{R_1}$ as the aggregate traffic arrival to VOQs or:
\begin{equation}
\label{lbc-voq-matrix-2a}
\mathbf{R_1} = \displaystyle\sum_{\gamma=0}^{t} \mathbf{A_1}(\gamma) 
\end{equation}
Let us assume the worse case scenario in (\ref{Dij_lbc}). Substituting (\ref{Dij_lbc}) into (\ref{eq:D1_lbc}), and (\ref{eq:D1_lbc}) and (\ref{lbc-voq-matrix-2a}) into (\ref{N_1(n)-2_lbc}), yields:
\begin{eqnarray}
\mathbf{N_1}(t) = \begin{dcases*}
\mathbf{N_1}(0) + \mathbf{R_1} - \mathbb{1}*\frac{t}{N} & for $\delta = 0$\\
\mathbf{N_1}(0) + \mathbf{R_1} - \mathbb{1}*\frac{t}{\delta N k} & for $\delta> 1$
\end{dcases*}
\label{N_1(n)-3_lbc}
\end{eqnarray}
From (\ref{N_1(n)-3_lbc}), we obtain:
\begin{eqnarray}
\begin{dcases*}
\lim_{t\to\infty}\frac{\mathbf{R_1}}{t} - \mathbb{1}*\frac{1}{N}  \leq \epsilon < \infty & for $\delta = 0$ \\
\lim_{t\to\infty}\frac{\mathbf{R_1}}{t} -  \mathbb{1}*\frac{1}{\delta N k} \leq \epsilon < \infty & for $\delta > 1$
\end{dcases*}
\label{VOQ_stab_lbc}
\end{eqnarray}
From the admissibility condition of $\mathbf{R_1}$, it is easy to see that for any value of $t$, (\ref{VOQ_stab_lbc}) is finite.
Hence, from the admissibility of $\mathbf{R_1}$, (\ref{N_1(n)-3_lbc}) and (\ref{VOQ_stab_lbc}), we conclude that occupancy of VOQ is weakly stable.

$\blacksquare$ \\

Now we prove VOMQs stability. As before, the queue occupancy matrix of VOMQs at time slot $t$ can be represented as:
\begin{equation}
\label{lbc-vomq-matrix-1}
\mathbf{N_2}(t) = \mathbf{N_2}(t-1) + \mathbf{A_2}(t) - \mathbf{D_2}(t)
\end{equation}
where $\mathbf{A_2}(t)$ is the arrival matrix at time slot $t$ to VOMQs and $\mathbf{D_2}(t)$ is the service rate matrix of VOMQs at time slot $t$. 
Solving (\ref{lbc-vomq-matrix-1}) recursively with consideration of an initial condition for $\mathbf{N_2}(t)$, yields:
\begin{equation}
\label{lbc-vomq-matrix-2}
\mathbf{N_2}(t) = \mathbf{N_2}(0) +  \displaystyle\sum_{\gamma=0}^{t} \mathbf{A_2}(\gamma) - \displaystyle\sum_{\gamma=0}^{t}\mathbf{D_2}(\gamma)
\end{equation}
Because a VOMQ is serviced at least once every $k$ time slots, the service rate of the VOMQ at $I_C(r,p)$ for $OP(v)$ at time slot $t$, $d_{2_{\mu,v}}(t)$ is:
\[d_{2_{\mu,v}}(t)= \frac{1}{k}~ \forall ~ \mu ~ and ~ v\]
Then, the service matrix of VOMQs is:
\begin{equation}\label{eq:D2_lbc}
\mathbf{D_2}(t)=[d_{2_{\mu,v}}(t)]
\end{equation}
and representing $\mathbf{R_2}$ as the aggregate traffic arrival to VOMQs or:
\begin{equation}
\label{lbc-vomq-matrix-2a}
\mathbf{R_2} = \displaystyle\sum_{\gamma=0}^{t} \mathbf{A_2}(\gamma) 
\end{equation} 
Substituting (\ref{eq:D2_lbc}) and (\ref{lbc-vomq-matrix-2a}) into (\ref{lbc-vomq-matrix-2}) gives:
\begin{equation}
\label{lbc-vomq-matrix-4}
\mathbf{N_2}(t) = \mathbf{N_2}(0) +  \mathbf{R_2} - \frac{1}{k} \mathbf{P_1}
\end{equation}
\begin{equation}
\mathbf{R_2} - \frac{1}{k} \mathbf{P_1} \leq \epsilon < \infty
\label{lbc-vomq-matrix-5}
\end{equation}
Recalling that $\mathbf{R_2}$ is admissible, per the discussion in Section III.A, and by substituting $\mathbf{P_1}$ and $\mathbf{R_2}$ into (\ref{lbc-vomq-matrix-5}), it is easy to see that $\epsilon$ is finite.
Hence, from (\ref{lbc-vomq-matrix-4}) and (\ref{lbc-vomq-matrix-5}), we conclude that the occupancy of VOMQ is weakly stable.

$\blacksquare$ \\

Now we prove the stability of CBs. The queue occupancy matrix of CBs at time slot $t$ can be represented as:
\begin{equation}
\label{lbc-cb-matrix-1}
\mathbf{N_3}(t) = \mathbf{N_3}(t-1) + \mathbf{A_3}(t) - \mathbf{D_3}(t)
\end{equation}
where $\mathbf{A_3}(t)$ is the packet arrival matrix at time slot $t$ CBs, and $\mathbf{D_3}(t)$ is the service rate matrix of CBs at time slot $t$. 
Solving (\ref{lbc-cb-matrix-1}) recursively as before yields:
\begin{equation}
\label{lbc-cb-matrix-2}
\mathbf{N_3}(t) = \mathbf{N_3}(0) +  \displaystyle\sum_{\gamma=0}^{t} \mathbf{A_3}(\gamma) - \displaystyle\sum_{\gamma=0}^{t}\mathbf{D_3}(\gamma)
\end{equation}
Because a CB is serviced at least once every $k$ time slots. Hence, the service rate of the CB at $OP(v)$ at time slot $t$, $d_{3_{v}}(t)$ is:
\[\frac{1}{k} \leq d_{3_{v}}(t) \leq 1\]
and service matrix of CBs is:
\begin{equation}\label{eq:D3_lbc}
\mathbf{D_3}(t)=[d_{3_{v}}(t)]
\end{equation}
Similarly, the aggregate traffic arrival to the CB or:
\begin{equation}
\label{lbc-cb-matrix-2a}
\mathbf{R_4} = \displaystyle\sum_{\gamma=0}^{t} A_3(\gamma) 
\end{equation}
Let us assume $d_{3_{v}}(t)= \frac{1}{k} ~ \forall ~v$ in (\ref{eq:D3_lbc}), which is the worst case scenario at which a CB gets served once every $k$ time slots. Substituting (\ref{eq:D3_lbc}) and (\ref{lbc-cb-matrix-2a}) into (\ref{lbc-cb-matrix-2}) gives:
\begin{equation}
\label{lbc-cb-matrix-3}
\mathbf{N_3}(t) = \mathbf{N_3}(0) + \mathbf{R_4} - \frac{1}{k} * \vec{1}
\end{equation}
where
\begin{equation}
\label{lbc-cb-matrix-4}
\mathbf{R_4} - \frac{1}{k} * \vec{1} \leq \epsilon < \infty
\end{equation}
With R4 being admissible, as discussed in Section III.A, and by substituting $\mathbf{R_4}$ into (\ref{lbc-cb-matrix-4}), it is easy to see that $\epsilon$ is finite.
Hence, from  (\ref{lbc-cb-matrix-3}) and (\ref{lbc-cb-matrix-4}), we conclude that the occupancy of CB is also weakly stable.\\

$\blacksquare$ \\

This completes the proof of Theorem \ref{theorem:throughput_lbc}.
\\
$\blacksquare$\color{black}

\bibliographystyle{IEEEtran}
\bibliography{Reference}

\end{document}